\documentclass[usegraphicx]{mn2e}

\usepackage{times, amssymb}

\title[Variability in red supergiant stars]{Variability in red supergiant stars:
pulsations, long secondary periods and convection noise}
\author[L.L. Kiss et al.]{L. L. Kiss$^1$, Gy. M. Szab\'o$^{2,3}$, T. R. Bedding$^1$\\
\\
$^1$School of Physics A28, University of Sydney, NSW 2006, Australia\\
$^2$Department of Experimental Physics and Astronomical Observatory, University of
Szeged, Hungary\\
$^3$Magyary Zolt\'an Postdoctoral Research Fellow}

\begin{document}

\date{Accepted ... Received ..; in original form ..}


\maketitle

\begin{abstract}

We study the brightness variations of galactic red supergiant stars using long-term
visual light curves collected by the American Association of Variable Star Observers
(AAVSO) over the last century. The full sample contains 48 red semiregular or irregular
variable stars, with a mean time-span of observations of 61 years. We determine periods
and period variability from analyses of power density spectra and time-frequency
distributions. We find two significant periods in 18 stars. Most of these periods fall
into two distinct groups, ranging from a few hundred to a few thousand days.
Theoretical models imply fundamental, first and possibly second overtone mode
pulsations for the shorter periods. Periods greater than 1000 days form a parallel
period-luminosity relation that is similar to the Long Secondary Periods of the 
Asymptotic Giant Branch stars. A number of individual power spectra shows a single mode
resolved into multiple peaks under a Lorentzian envelope, which we interpret as
evidence for stochastic oscillations, presumably caused by the interplay of convection
and pulsations. We find a strong $1/f$ noise component in the power spectra that is
remarkably similar in almost all stars of the sample. This behaviour fits the picture
of irregular photometric variability caused by large convection cells, analogous to the
granulation background seen in the Sun.

\end{abstract}


\section{Introduction}

Red supergiants (RSGs) are evolved, moderately massive (10--30 M$_\odot$) 
He-burning stars. As such, they are key agents of nucleosynthesis and chemical
evolution of the Galaxy, and have also long been known for their slow optical
variations. This variability is usually attributed to radial pulsations
(Stothers 1969, Wood, Bessell \& Fox 1983, Heger et al.\ 1997, Guo \& Li 2002), although
irregular variability caused by huge convection cells was also suggested from
theory (Schwarzschild 1975, Antia et al.\ 1984) and observations (e.g. Tuthill, Haniff \&
Baldwin 1997). The presence of oscillations, in principle, offers the possibility
of asteroseismology, i.e. measuring frequencies of stars and then
identifying the modes of pulsations via comparing the observations to theoretical
models. However, the time scales involved in RSGs makes this approach very
difficult. Model calculations predict pulsational
instability for fundamental and low-order overtone modes (Stothers 1972, Wood,
Bessell \& Fox 1983, Heger et al.\ 1997, Guo \& Li 2002), with fundamental periods ranging
from 150 to 4,000 days. Consequently, any meaningful period determination may 
require decades of observations. Here we analyse a homogeneous sample of RSG light
curves with a typical time-span of over 22,000 days, which is the longest available
observational data for these objects.

Significant interest in RSG pulsations was driven by the discovery of their distinct
period-luminosity (P--L) relation. Although the relation is not as tight as for
Cepheids or red giant stars, various authors pointed out that because of their great
intrinsic brightnesses, RSGs might be useful as  extragalactic distance indicators
(e.g. Glass 1979, Feast et al.\ 1980, Wood \& Bessell 1985). A recent example is  that
of Pierce et al.\ (2000), who re-calibrated the RSG P--L relation in the  near-infrared
and then measured the distance to M101 from 42 RSGs in that galaxy  (Jurcevic et al.
2000). In this respect, galactic RSGs that are members of young O--B associations are
particularly important because they can serve as local calibrators of the P--L
relation -- provided that reliable periods can be measured. Establishing this zero
point is one of the by-products of this paper.

Until recently, there was a serious discrepancy between evolutionary models and the
empirical Hertzsprung--Russell diagram of RSGs. In particular, stellar evolutionary
models seemed unable to produce RSGs that are as cool or as luminous as observed
(Massey 2003, Massey \& Olsen 2003). A solution was found by Levesque et al.\ (2005),
who derived a new (hotter) temperature scale from optical spectrophotometry for 74
Galactic RSGs, so that most of the discrepancy was removed. In addition, Massey et al.
(2005) argued that circumstellar dust around RSGs may account for many magnitudes of
extra extinction compared to other stars in the same O--B associations, so that
determining fundamental  physical parameters can be quite difficult even for the
brightest RSGs. The problem is very well illustrated, for instance, by the peculiar
RSG VY~CMa, for which previous radius estimates of up to 2,800 R$_\odot$ (Smith et al.
2001) were  scaled down to 600 R$_\odot$ (Massey, Levesque \& Plez 2006), due to the
hotter temperature scale alone. These wildly different parameter values could be 
tested if pulsation mode identification was made possible, which is another motivation
for this work.

In the General Catalogue of Variable Stars (GCVS, Kholopov et al.\ 1985-1988) there are
two categories of variable RSGs: SRc and Lc. SRc type stars are semiregular  
late-type supergiants with amplitudes of about 1 mag and periods from 30 days  to
several thousand days. In contrast, Lc type stars are irregular variable supergiants
having visual amplitudes of about 1 mag. As the distinction between {\it semiregular}
and {\it irregular} behaviour is not well defined in the literature, stars in both
categories can possibly reveal important information. So far, the most detailed study
of bright galactic RSG variables as a group was that by  Stothers \& Leung (1971), who
discussed periods, luminosities and masses for  22 stars. Their periods were either
taken from papers published in the 1950s or (for 8 stars) determined by the authors
using visual data. For some stars, such as  $\alpha$~Ori, $\alpha$~Her and $\mu$~Cep,
one can find more recent attempts to derive period(s) using various sources of
photometric or radial velocity data  (e.g. Mantegazza 1982, Smith, Patten \& Goldberg
1989, Percy et al.\ 1996, Brelstaff et al.\ 1997,  Rinehart et al.\ 2000). However, the
typical time-span of analysed observations rarely exceeded 10--15 years and usually
with many gaps, thus the derived periods (or mean cycle lengths) did not always agree.

The American Association of Variable Star Observers (AAVSO) has been collecting visual
observations of variable stars for almost a century. The AAVSO database is by far the
biggest one of its kind, containing over 12 million individual brightness estimates
for about 6,000 variable stars. Of these, roughly 50 stars are reasonably
well-observed bright galactic RSGs of the GCVS types SRc and Lc. Using their
observational records in the AAVSO database, we have carried out a period and light
curve analysis of this group of stars. The main aim was to derive periods for all
variables. However, it turned out that in some cases the light curves had long enough
time-span to study the long-term behaviour of their brightness fluctuations.
Therefore, in addition to determining the dominant time-scales, we also discuss the
nature and extent of the irregularities.

The paper is organised as follows. In Sect. 2 we describe the sample selection,
data processing and analysis. Sect. 3 contains the results, the newly determined periods;
Sect. 4 discusses the multiperiodic nature and a comparison with pulsation model
predictions. In Sect. 5 we present evidence for stochastic oscillations in some of
the stars and the presence of a strong $1/f$ noise. Sect. 6 briefly summarizes the
main findings of this study.

\section{Sample selection and data analysis}

\begin{figure*}
\begin{center}  
\leavevmode
\includegraphics[width=14cm]{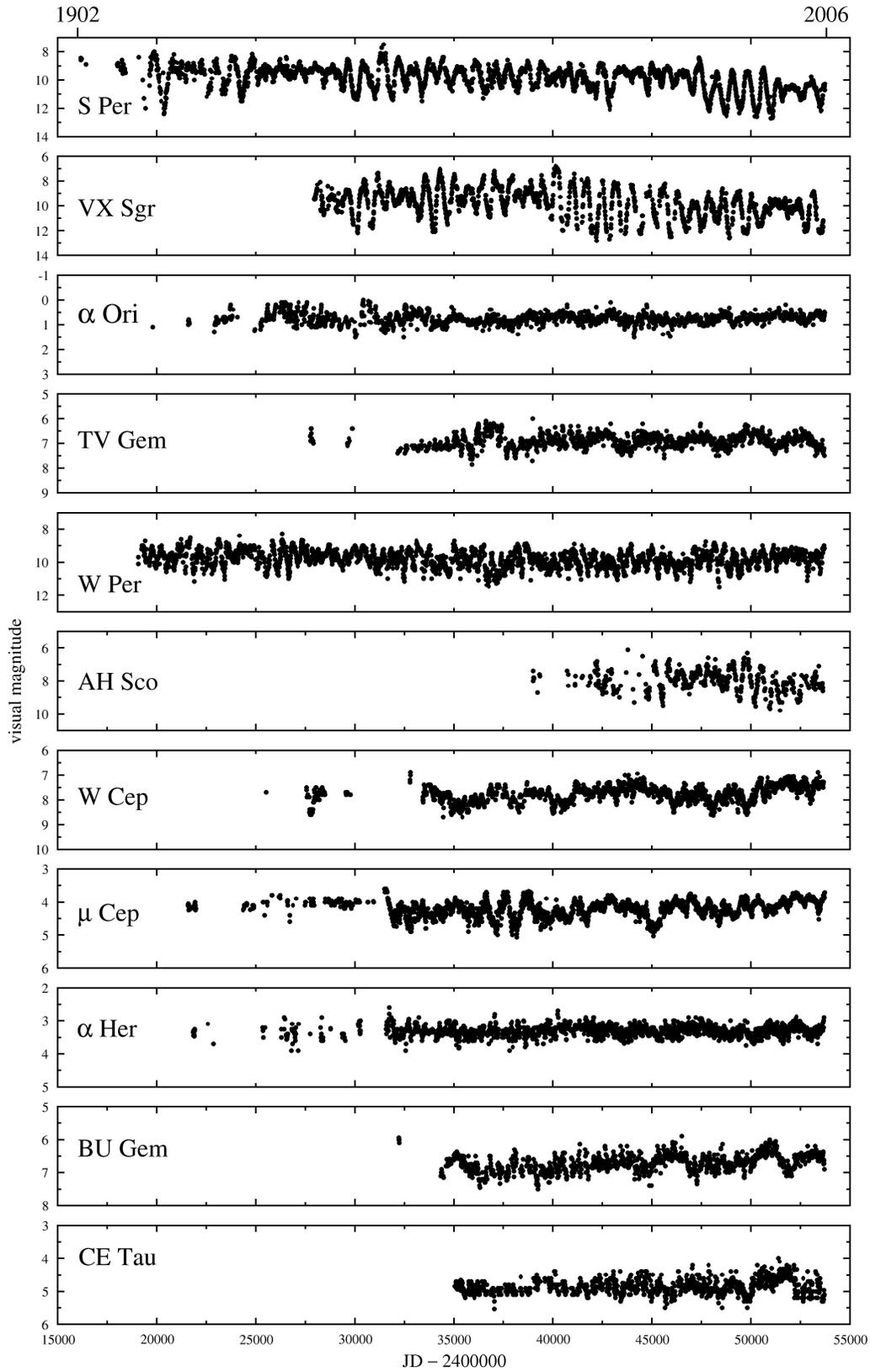}
\end{center}
\caption{Sample AAVSO light curves of red supergiant variables (10-day bins). Note
the different magnitude scale in each plot.}
\label{srclc}
\end{figure*}

\begin{table*}
\begin{centering}
\caption{\label{sample} The basic properties of the sample. Classification and
spectral types were taken from the General Catalog of Variable Stars; m$_{\rm max}$
and m$_{\rm min}$ are the maximum and minimum visual magnitudes as measured from
the binned light curves. Every dataset ends in late 2005.}
\begin{tabular}{lcllrrrrr}
\hline
Star & IRAS-identifier & Class & Sp. type & m$_{\rm max}$ & m$_{\rm min}$ &JD [yr] start
& JD end & No. points\\
\hline
SS~And & 23092+5236 & SRc & M6II & 9.0 & 10.1 & 2429543 [1939] & 2453735 & 445\\
NO~Aur & 05374+3153 & Lc & M2sIab & 6.0 & 6.5 & 39740 [1967]& 53488 & 425\\
UZ~CMa & 06165--1701 & SRc & M6II & 11.0 & 12.0 & 39385 [1966] & 53700 & 1544\\
VY~CMa & 07209--2540 & * & M5eIbp & 7.4 & 9.8 & 36168 [1957] & 53705 & 5247\\
RT~Car & 10428--5909 & Lc & M2Ia & 8.3 & 9.5 & 39159 [1965] & 53730 & 1324\\
BO~Car & -- & Lc & M4Ib & 7.0 & 8.0 & 39159 [1965] & 53729& 1226\\
CK~Car & 10226--5956 & SRc & M3.5Iab & 7.2 & 8.5 & 39156 [1965] & 53357 & 737\\
CL~Car & 10520--6049 & SRc & M5Iab & 8.2 & 9.6 & 39153 [1965] & 53686 & 539\\
EV~Car & 10186--6012 & SRc & M4.5Ia & 7.4 & 9.0 & 39154 [1965] & 53357 & 694\\
IX~Car & 10484--5943 & SRc & M2Iab & 7.2 & 8.5 & 39165 [1965] & 53730 & 1069\\
TZ~Cas & 23504+6043 & Lc & M2Iab & 8.9 & 10.1 & 34992 [1954] & 53705 & 826\\
PZ~Cas & 23416+6130 & SRc & M2-M4Ia & 8.2 & 10.2 & 40153 [1968] & 53675 & 990\\
W~Cep & 22345+5809 & SRc & M2epIa & 7.0 & 8.5 & 25528 [1929] & 53700 & 6987\\
ST~Cep & 22282+5644 & Lc & M2Iab & 7.9 & 8.9 & 27838 [1934] & 53679 & 1242\\
$\mu$~Cep & 21419+5832 & SRc & M2eIa & 3.7 & 4.9 & 21566 [1917] & 53732 & 40044\\
T~Cet & 00192--2020 & SRc & M5-M6IIe & 5.2 & 6.8 & 20177 [1913] & 53724 & 6449\\
AO~Cru & 12150--6320 & Lc & M0Iab & 7.5 & 8.3 & 38895 [1964]& 53686 & 1396\\
RW~Cyg & 20270+3948 & SRc & M2-M4Iab & 8.0 & 9.5 & 16428 [1903] & 53693 & 1871\\
AZ~Cyg & -- & SRc & M2-4Iab & 7.8 & 10.0 & 29451 [1939] & 53684 & 574\\
BC~Cyg & 20197+3722 & Lc & M3Iab: & 9.0 & 10.8 & 33206 [1949] & 53736 & 777\\
BI~Cyg & 20194+3646 & Lc & M4Iab & 8.6 & 10.6 & 33264 [1949] & 53736 & 1002\\
TV~Gem & 06088+2152 & SRc & M1.3Iab & 6.3 & 7.4 & 27750 [1934] & 53704 & 8449\\
WY~Gem & 06089+2313 & Lc+E: & M2epIab & 7.2 & 7.7 & 25256 [1927] & 53706 & 5548\\
BU~Gem & 06092+2255 & Lc & M1-M2Iab & 6.1 & 7.2 & 32213 [1946] & 53716 & 7598\\
IS~Gem & 06464+3239 & SRc & K3II & 5.6 & 6.0 & 40312 [1969] & 53731 & 2758\\
$\alpha$~Her & -- & SRc  &  M5Iab: & 3.0 & 3.6 & 21840 [1918] & 53708 & 13853\\
RV~Hya & 08372--0924 & SRc & M5II & 7.4 & 8.1 & 21662 [1917] & 51634 & 416\\
W~Ind & 21108--5314 & SRc & M4-M5IIe & 8.5 & 10.2 &  42171 [1973] & 52937 & 399\\
Y~Lyn & 07245+4605 & SRc & M6sIb & 6.8 & 8.2 & 34783 [1953] & 53730 & 7992\\
XY~Lyr & 18364+3937 & Lc & M4-M5Ib & 5.7 & 6.6 & 32500 [1947] & 53709 & 6933\\
$\alpha$~Ori & 05524+0723 & SRc & M2Iab: & 0.3 & 1.2 & 21597 [1917] & 53731 & 19976\\ 
S~Per & 02192+5821 & SRc & M3Iae & 8.1 & 12.6 &  16160 [1902] & 53734 & 24863\\
T~Per & 02157+5843 & SRc & M2Iab & 8.5 & 9.2 & 16160 [1902] & 53711 & 8726\\
W~Per & 02469+5646 & SRc & M3Iab & 8.7 & 11.3 & 19069 [1910] & 53728 & 16687\\
RS~Per & 02188+5652 & SRc & M4Iab & 8.0 & 9.8 & 33490 [1950] & 53700 & 2045\\
SU~Per & 02185+5622 & SRc & M3.5Iab & 7.3 & 8.7 & 21396 [1915] & 53700 & 3121\\
XX~Per & 01597+5459 & SRc  & M4Ib & 7.9 & 9.0 & 23469 [1922] & 53709 & 3044\\
AD~Per & 02169+5645 & SRc & M3Iab & 7.5 & 9.0 & 29545 [1939] & 53700 & 3038\\
BU~Per & 02153+5711 & SRc & M3.5Ib & 8.5 & 10.0 & 33572 [1950] & 53697 & 1413\\
FZ~Per & 02174+5655 & SRc & M0.5-M2Iab & 7.8 & 8.7 & 33897 [1950] & 53697 & 1875\\
KK~Per & 02068+5619 & Lc & M1-M3.5Iab & 7.7 & 8.4 & 40689 [1970] & 53697 & 1680\\
PP~Per & 02135+5817 & Lc & M0-M1.5Iab & 9.0 & 9.5 & 42041 [1973] & 53697 & 1382\\
PR~Per & 02181+5738 & Lc & M1Iab & 7.6 & 8.4 & 40689 [1970] & 53697 & 1616\\
VX~Sgr & 18050--2213 & SRc & M4e-M10eIa & 6.9 & 12.7 & 27948 [1934] & 53664 & 6300\\
AH~Sco & 17080--3215 & SRc & M4III: & 6.5 & 9.6 & 38994 [1965] & 53668 & 1262\\
$\alpha$~Sco & 16262--2619 & Lc & M1.5Iab-b & 0.6 & 1.6 & 21457 [1916] & 53584 & 462\\
CE~Tau & 05292+1833 & SRc & M2Iab & 4.3 & 5.2 & 35053 [1954] & 53723 & 2767\\
W~Tri & 02384+3418 & SRc & M5II & 7.6 & 8.8 &  31018 [1943] & 53705 & 4916\\
\hline
\end{tabular}
\end{centering}
\end{table*}

To find all well-observed RSG variables, we browsed through the AAVSO light curves  of
all stars listed as SRc or Lc type red supergiant in the GCVS. As an independent
source of well-known galactic RSG stars, we checked the objects in Stothers \& Leung
(1972), Pierce et al.\ (2000) and Levesque et al.\ (2005). The latter paper was also
used to collect the main physical parameters of the sample. The full set of stars is
presented in Table\ \ref{sample} where, in addition to the GCVS data, we also indicate
the starting and ending Julian Dates of the AAVSO light curves and the number of
points in the raw data. We kept stars with at least 400 individual points (actually,
W~Ind only has 399). In total, we retained 234,527 visual magnitude estimates for 48
stars, with a total time-span of 2,926 years, or about 61 years of data for
each star.

The data were handled in a similar way to our previous analyses of long-term
visual light curves (e.g. Kiss et al.\ 1999, 2000; Kiss \& Szatm\'ary 2002; Bedding et
al.\ 2005). The raw light curves were plotted and then inspected for outlying points,
which were removed by a sigma-clipping procedure. 10-day bins  were calculated, which
helped make the data distribution more even because in every star, there was
gradual increase in the frequency of observations by a factor of 3 to 5 over the 20th
century. Without binning, any kind of period determination would have been strongly
biased by the latter half of the data. 

Representative light curves are shown in Fig.\ \ref{srclc}, where the time axes have
the same range to give a comparative illustration of the time-spans of the data. We
see various characteristic features in the light curves. The most impressive stars are
those with the largest amplitudes, like S~Per and  VX~Sgr, where the full brightness
range is similar to that of a Mira-type variable. However, none of the stars are as 
regular as a Mira. At the other end of the spectrum, we see stars with low amplitudes
($<$1 mag full range) and in many cases, two separate time-scales of variations: a
slow one of a few thousand days and a faster one of a few hundred days (e.g.
$\alpha$~Ori, TV~Gem, $\alpha$~Her). In between, we also see intermediate-amplitude
objects (of 2 to 3 mags), whose light curves are dominated by relatively stable
cycles, even suggesting the presence of stable  oscillations (W~Per, AH~Sco). However,
in many of the stars, there seems to be no stable periodicity, although the data
clearly show real brightness fluctuations (e.g. $\mu$~Cep, BU~Gem, CE~Tau).   

\begin{figure*}
\begin{center}  
\leavevmode
\includegraphics[width=8.5cm]{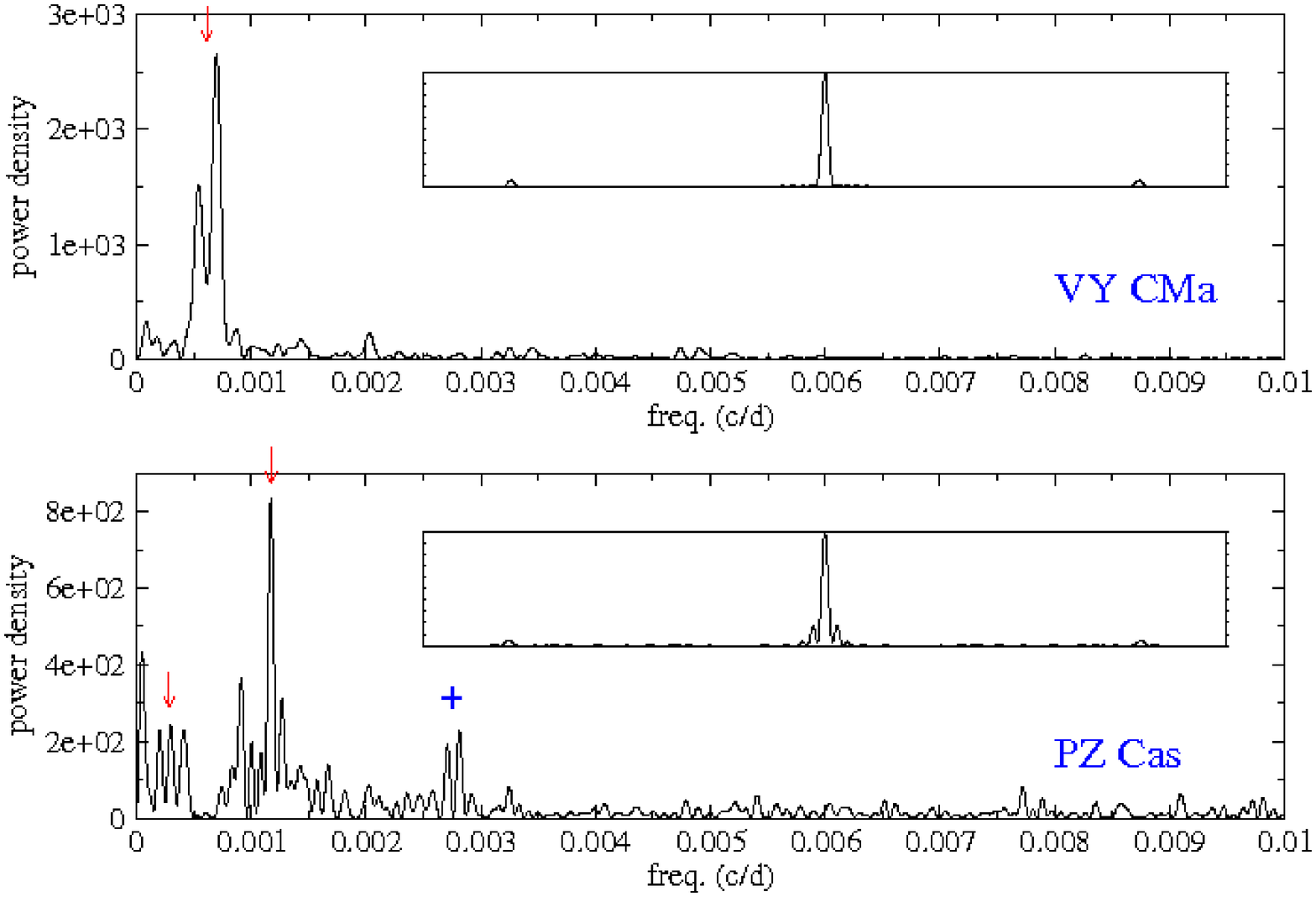}
\includegraphics[width=8.5cm]{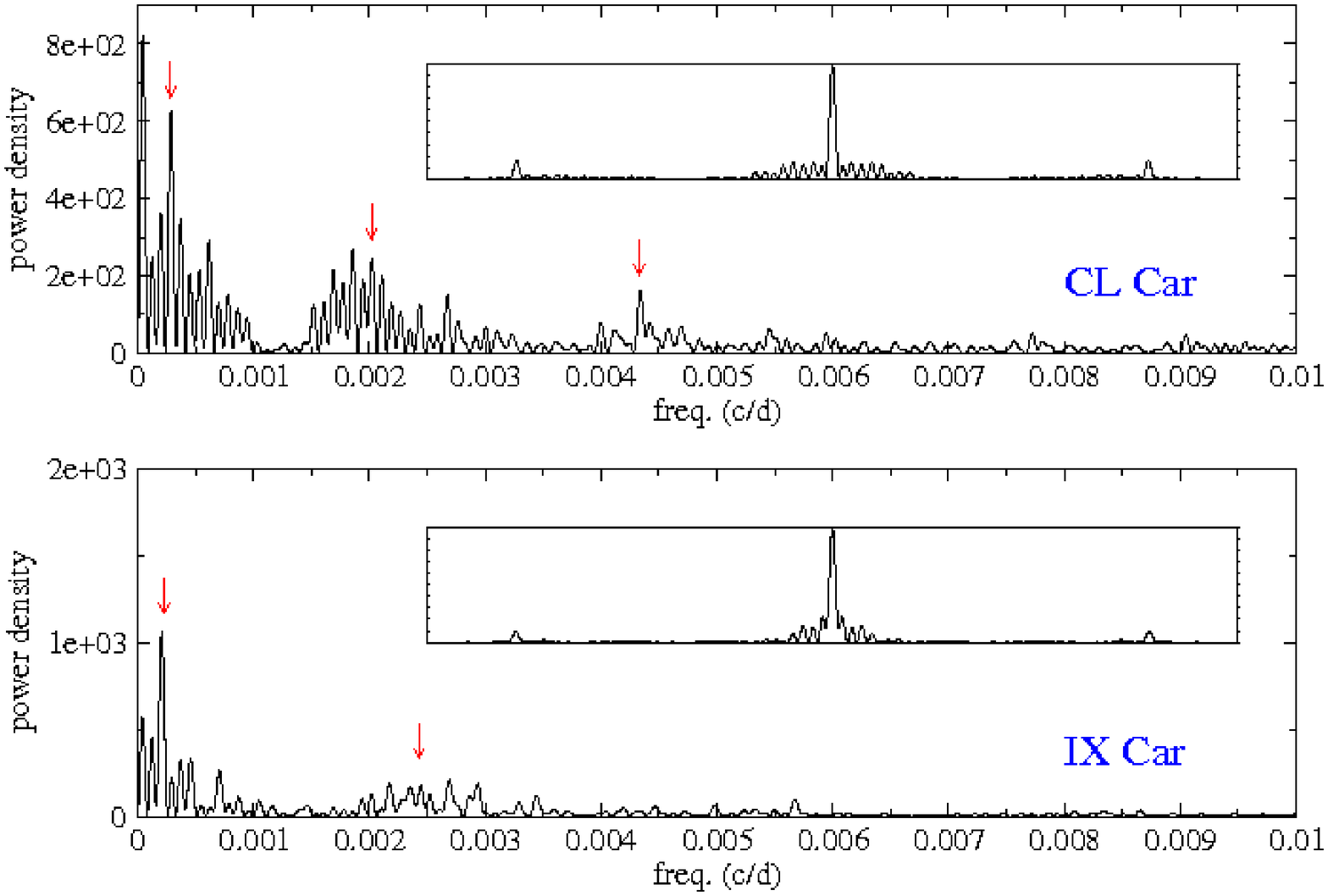}
\vskip3mm
\includegraphics[width=8.5cm]{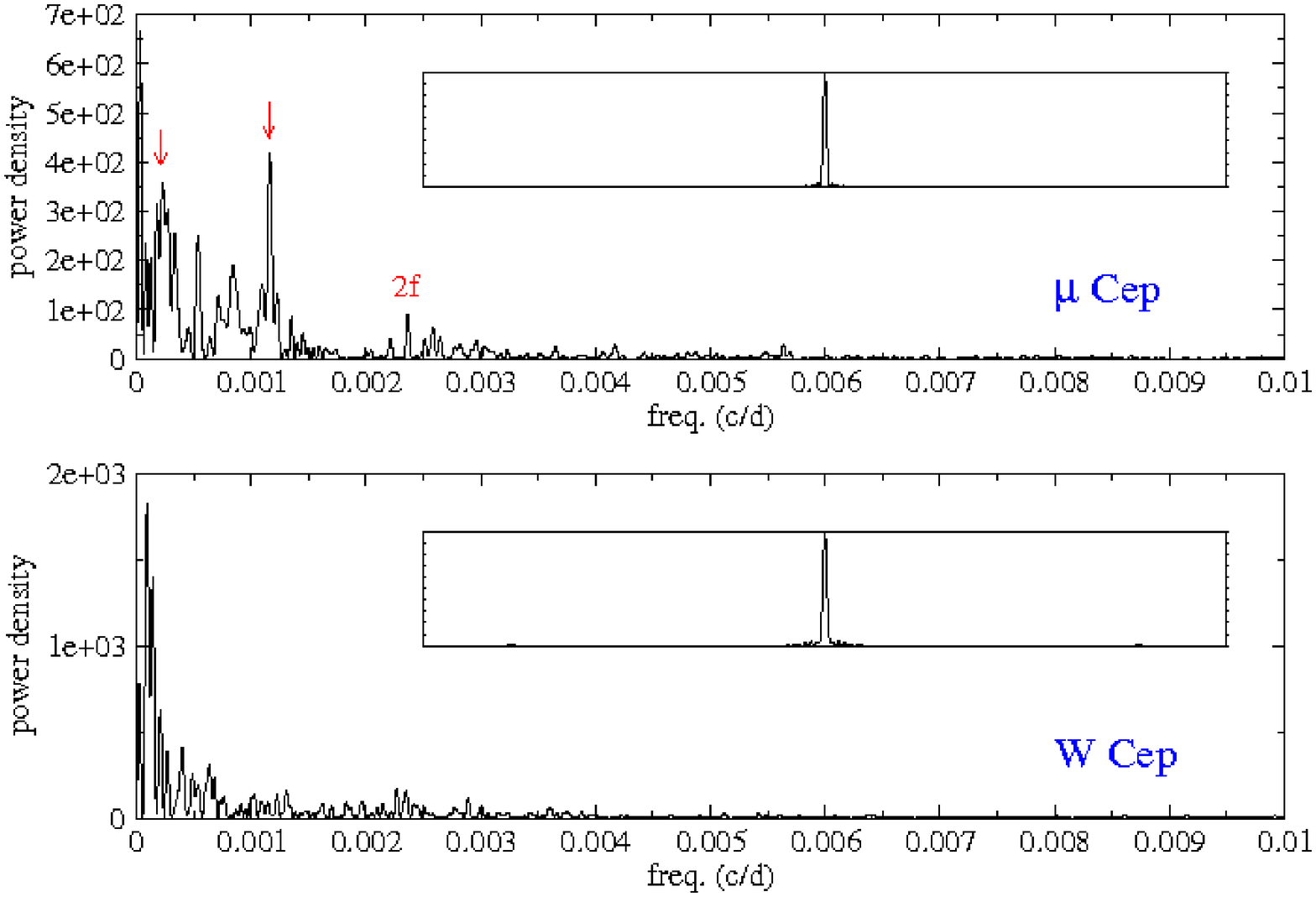}
\includegraphics[width=8.5cm]{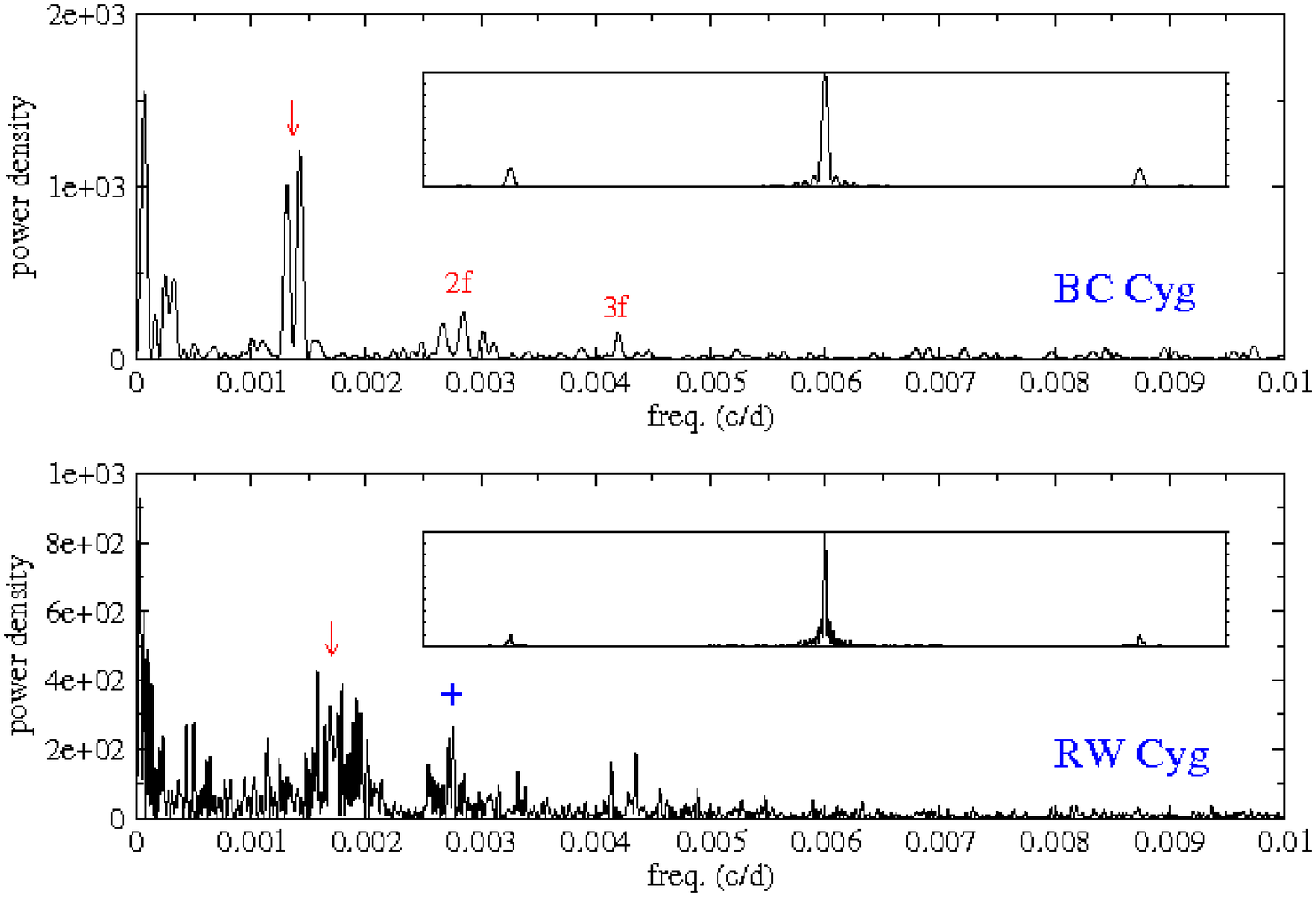}
\vskip3mm
\includegraphics[width=8.5cm]{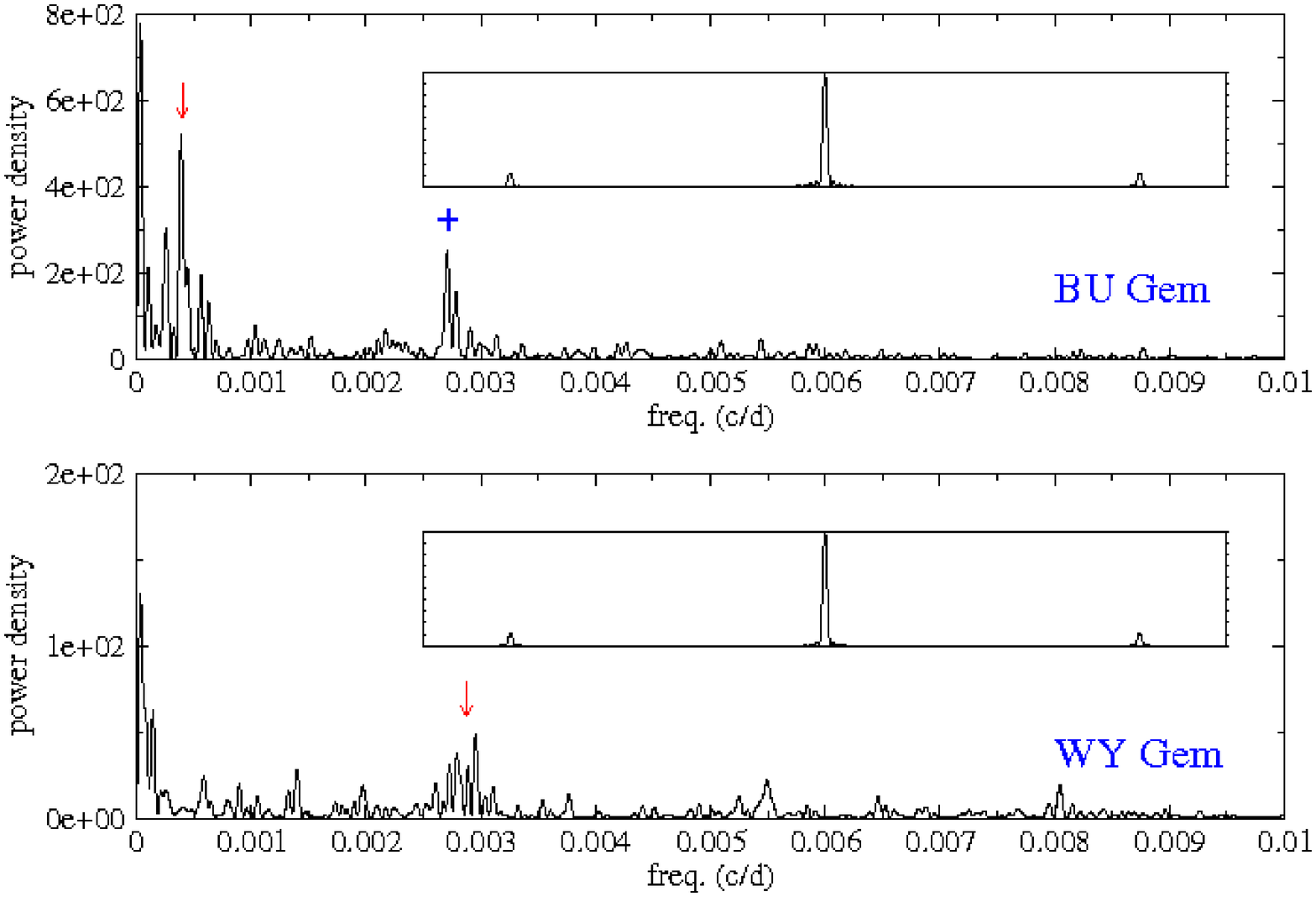}
\includegraphics[width=8.5cm]{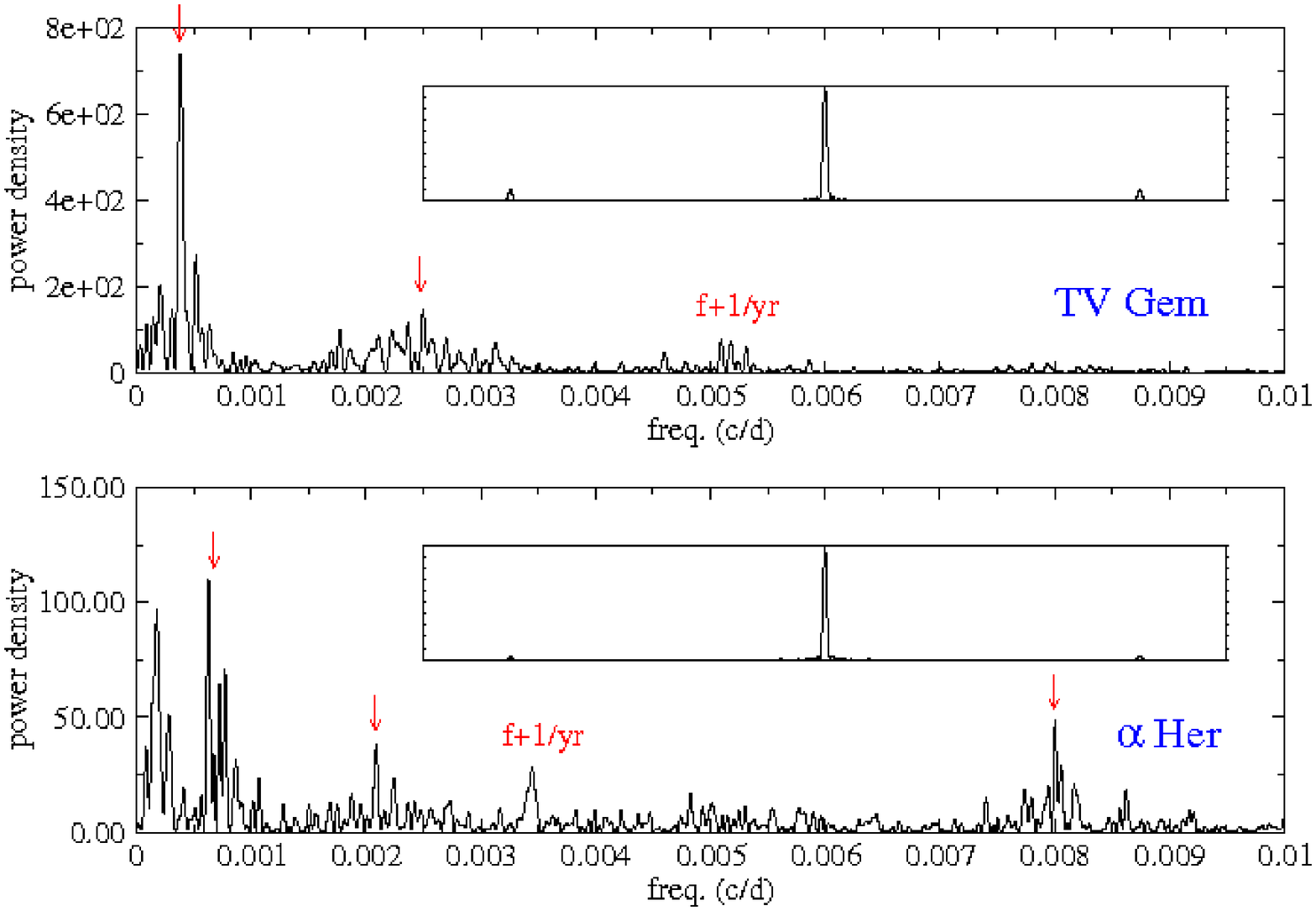}
\end{center}
\caption{Power density spectra of AAVSO data. The insets show the spectral window
on the same scale, while the arrows mark the adopted frequencies. Integer 
multiples  of a main frequency ($2f$, $3f$, etc.), yearly aliases ($f$+1/yr) 
or exactly 1-yr periods (+ signs) were not treated as separate periods.}
\label{pds1}
\end{figure*}

\begin{figure*}
\begin{center}  
\leavevmode
\includegraphics[width=8.5cm]{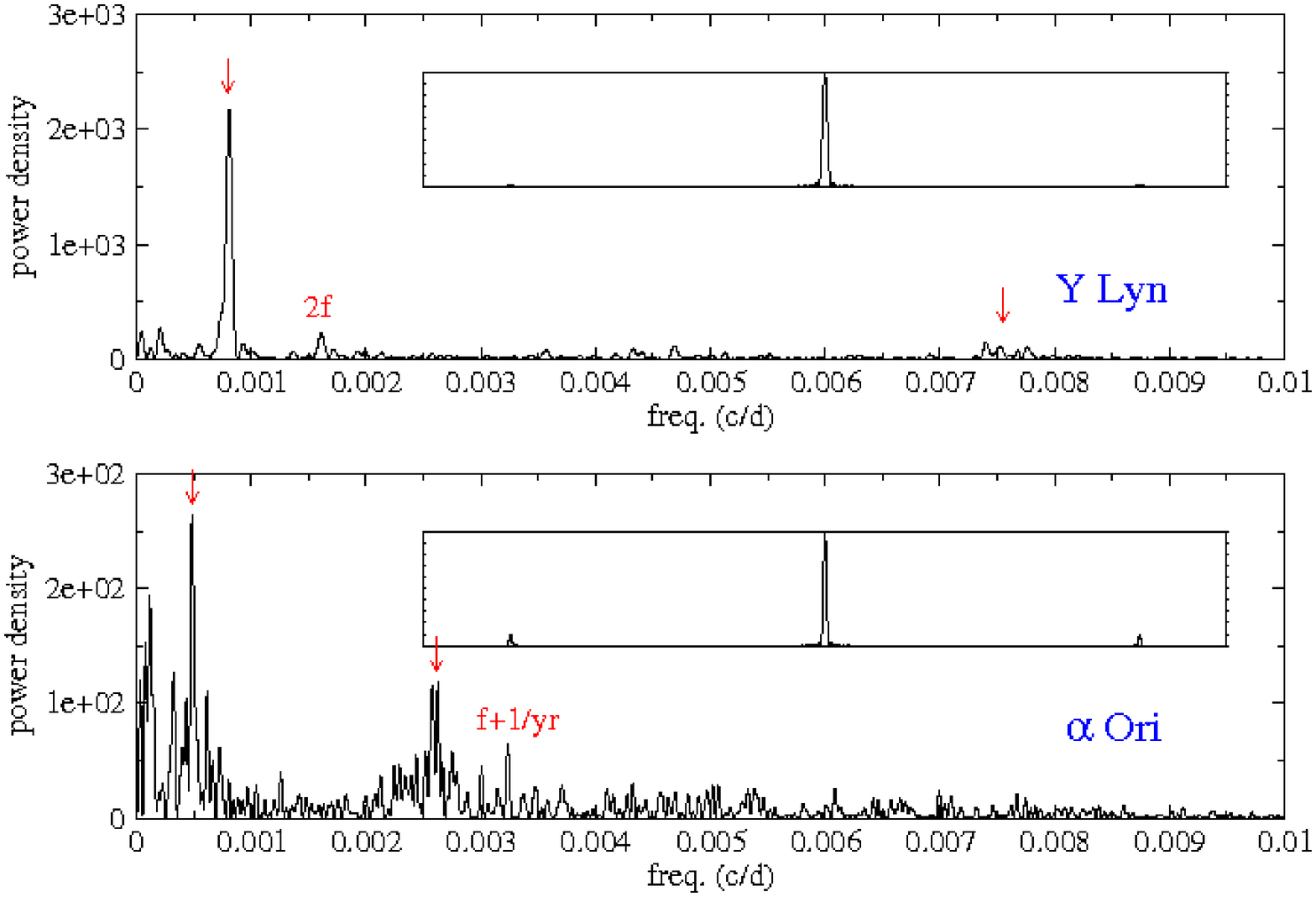}
\includegraphics[width=8.5cm]{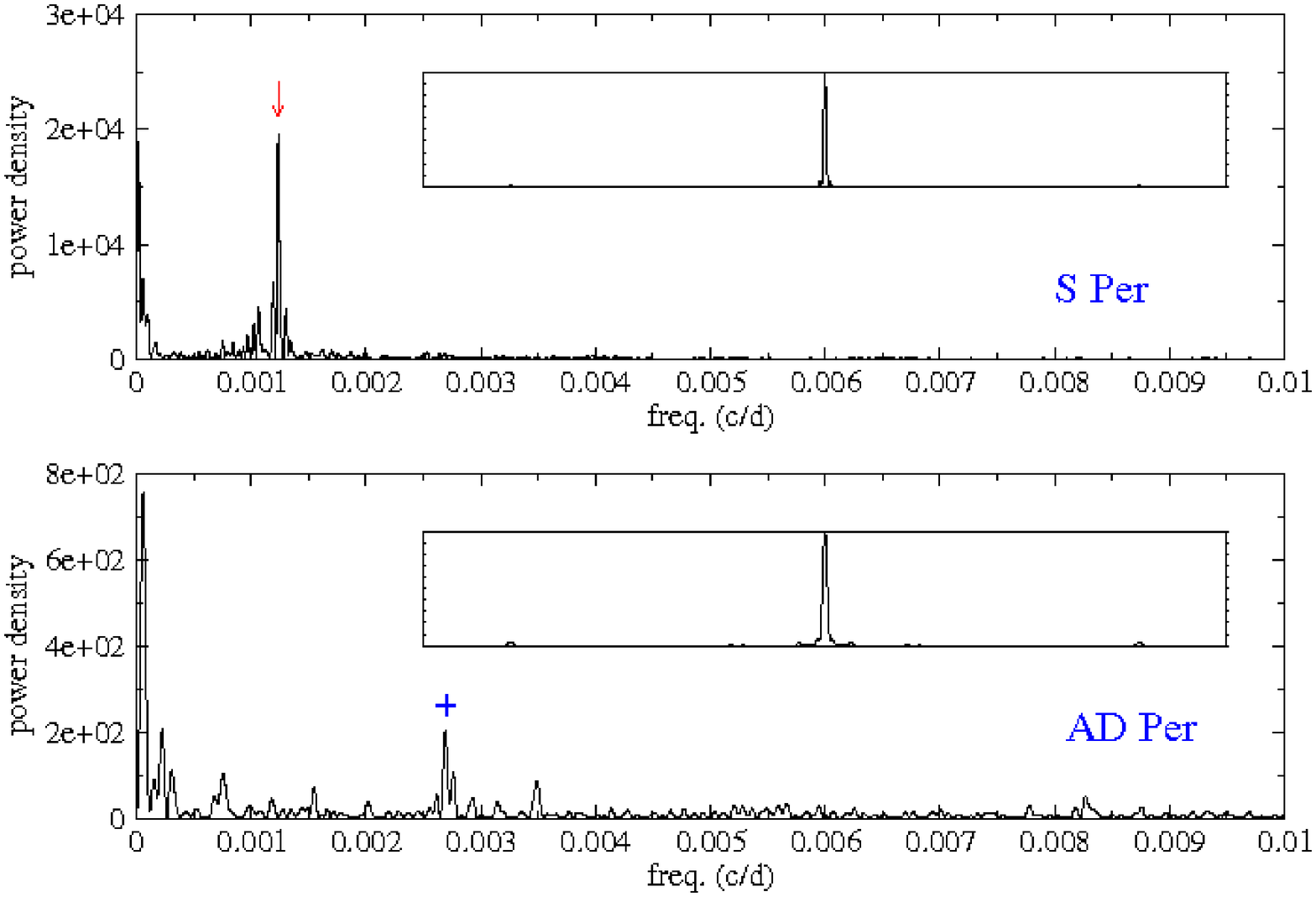}
\vskip3mm
\includegraphics[width=8.5cm]{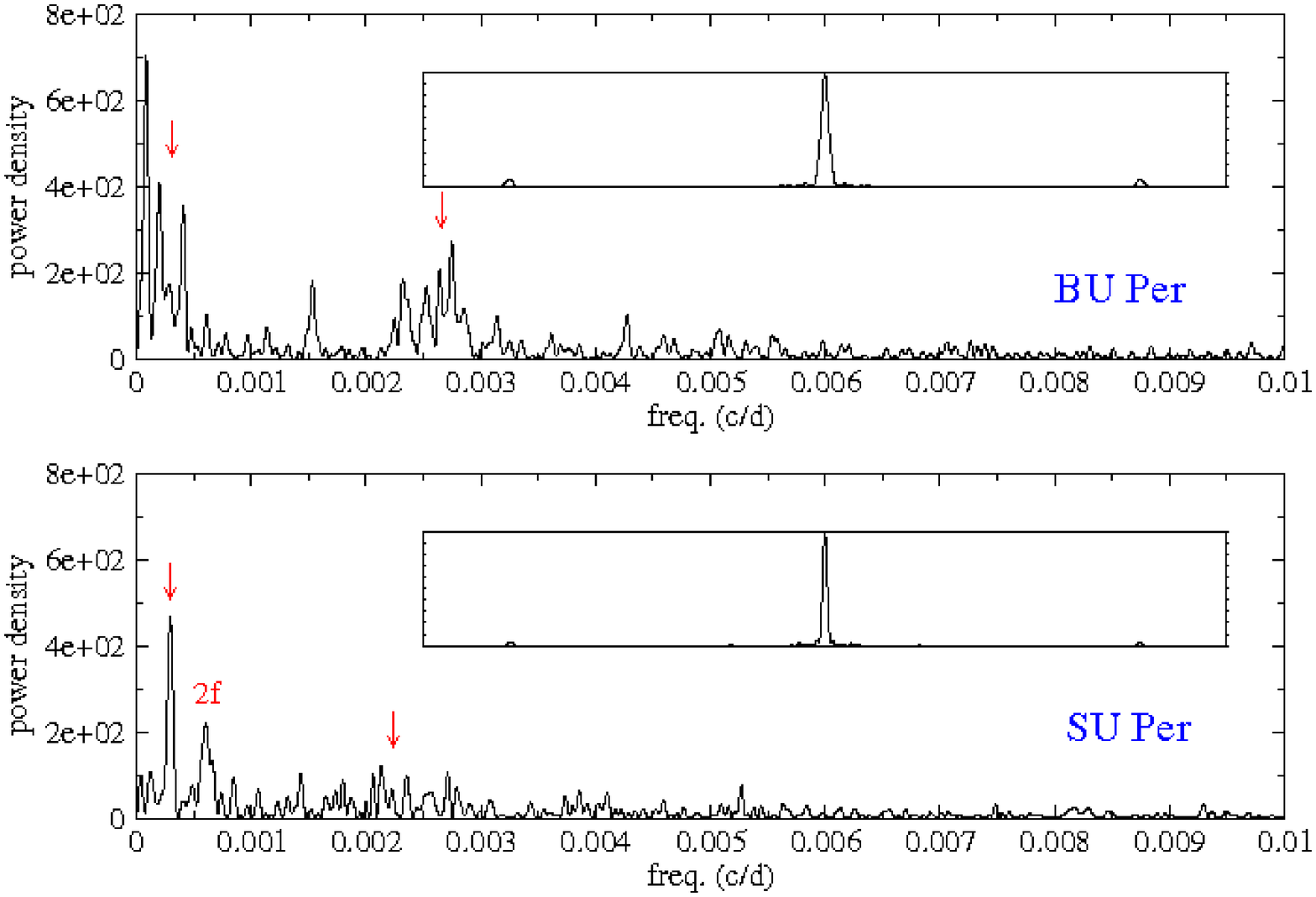}
\includegraphics[width=8.5cm]{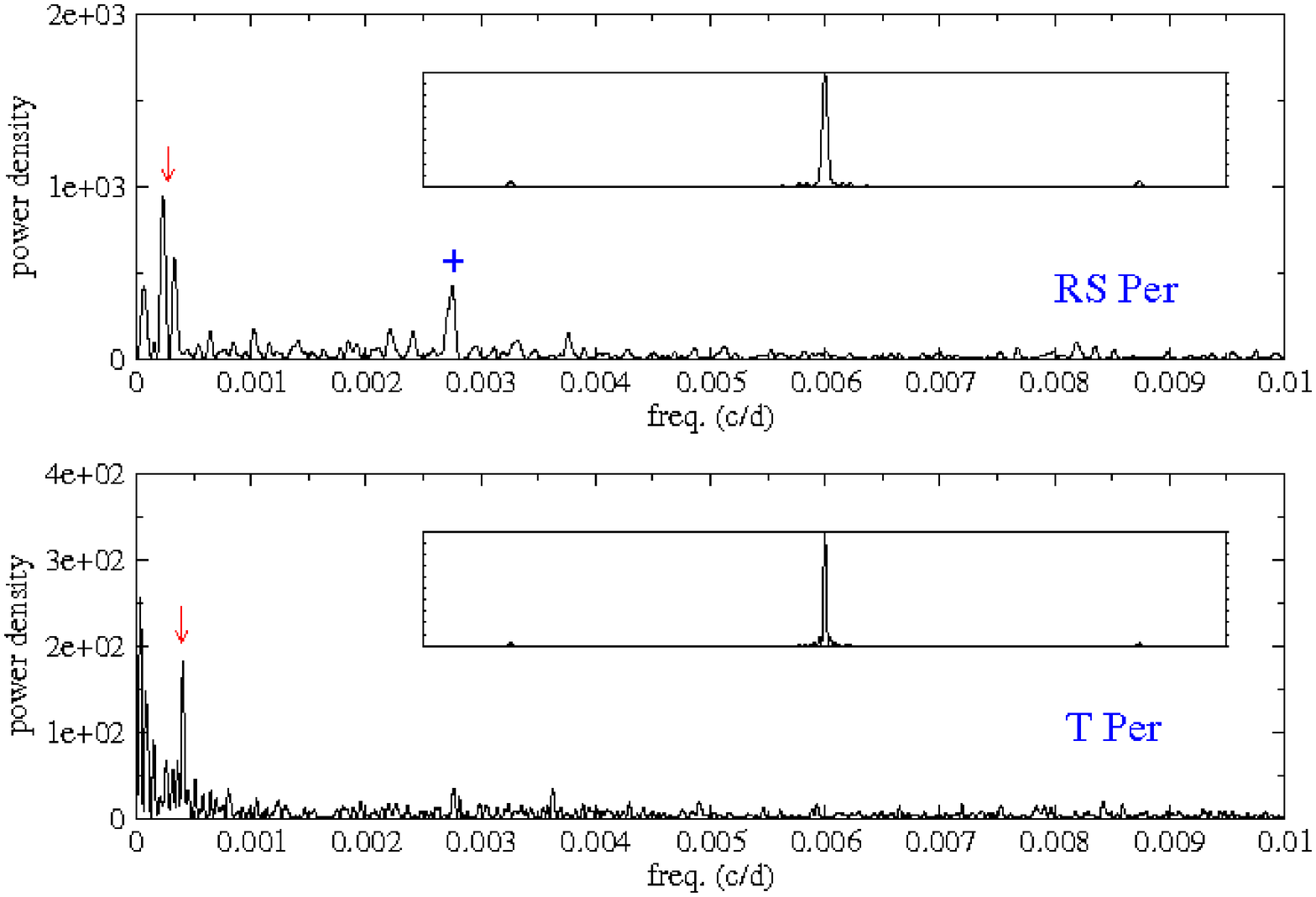}
\vskip3mm
\includegraphics[width=8.5cm]{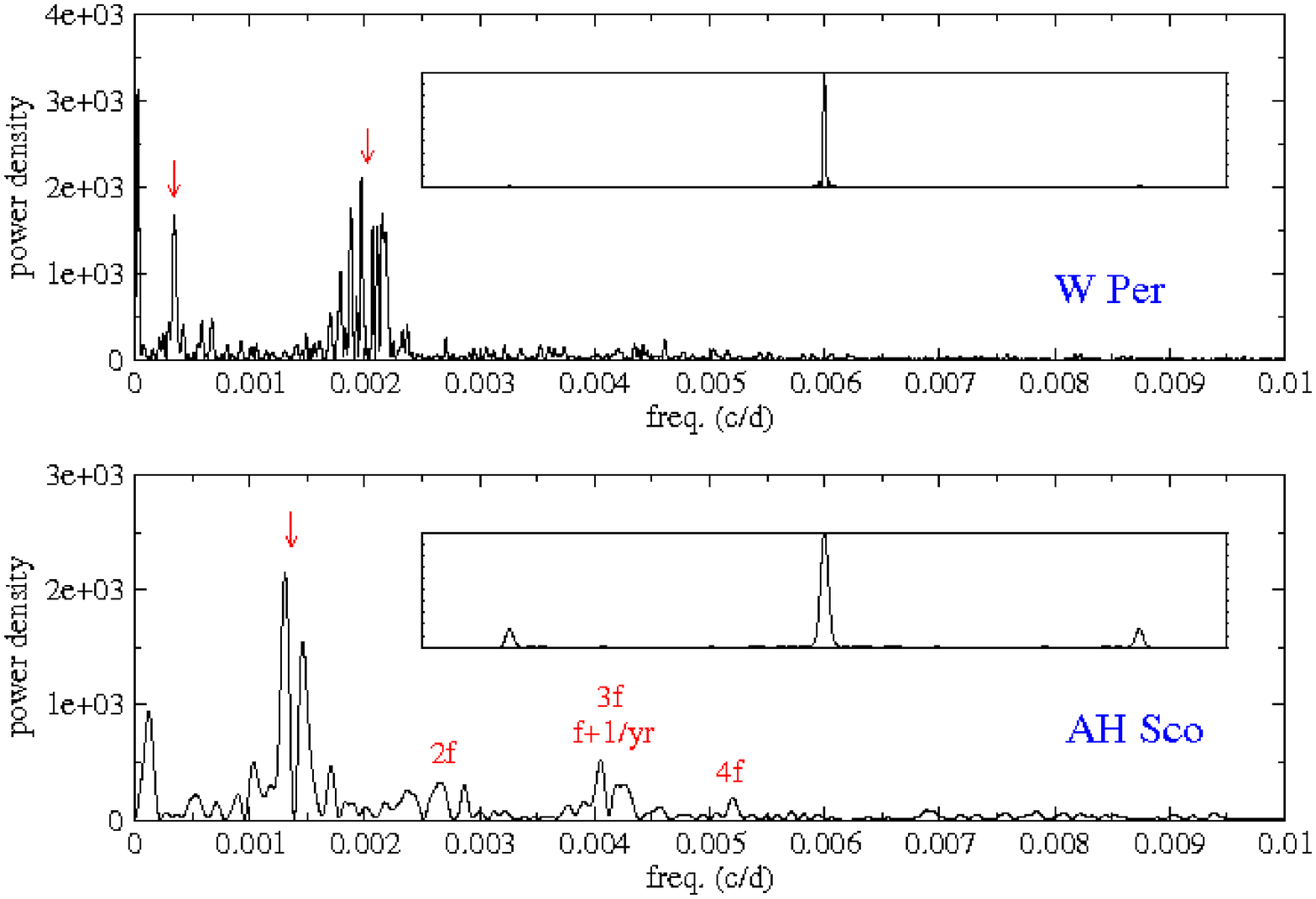}
\includegraphics[width=8.5cm]{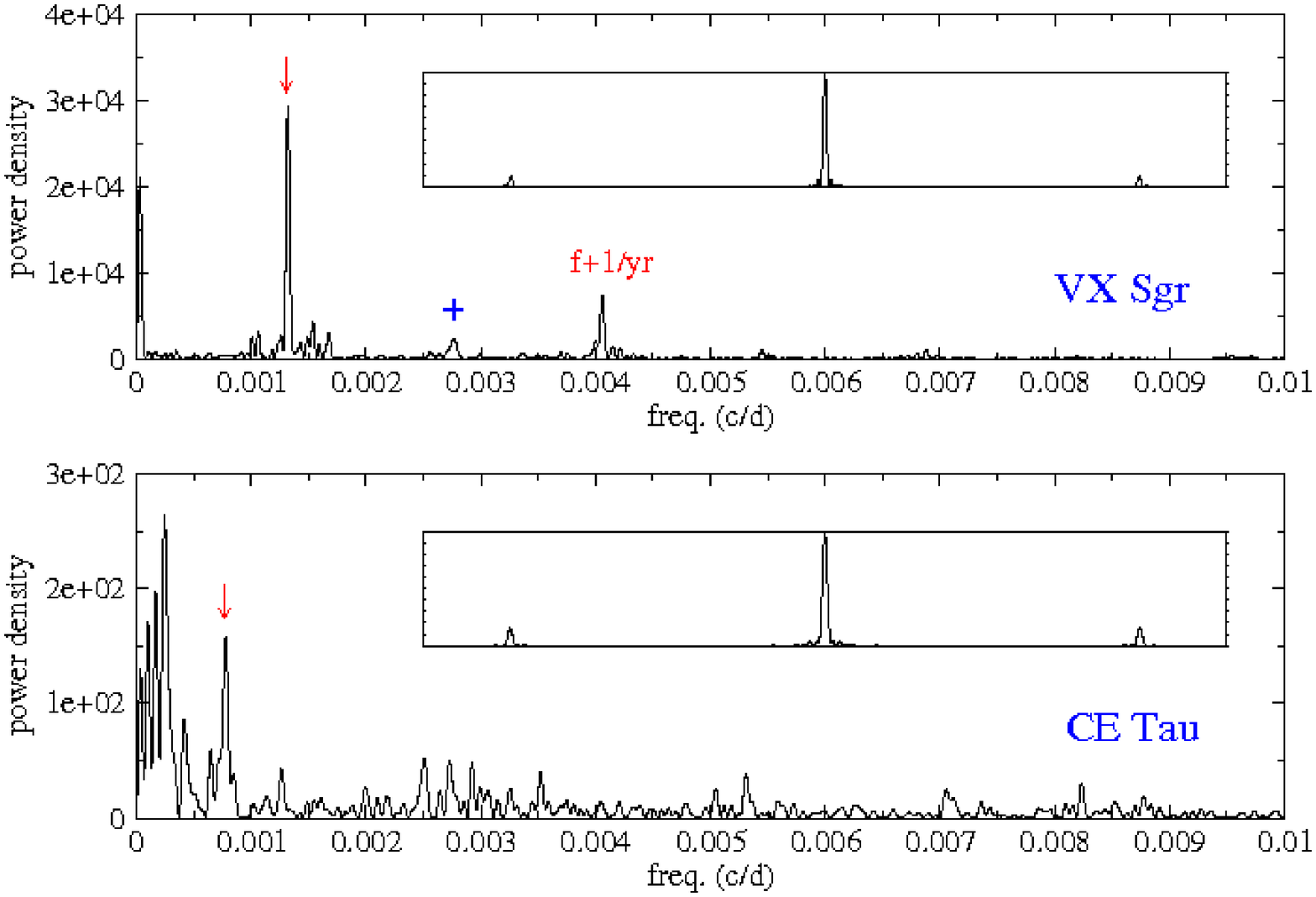}
\end{center}
\caption{Sample power density spectra (cont.)}
\label{pds2}
\end{figure*}

\begin{figure}
\begin{center}  
\leavevmode
\includegraphics[width=8cm]{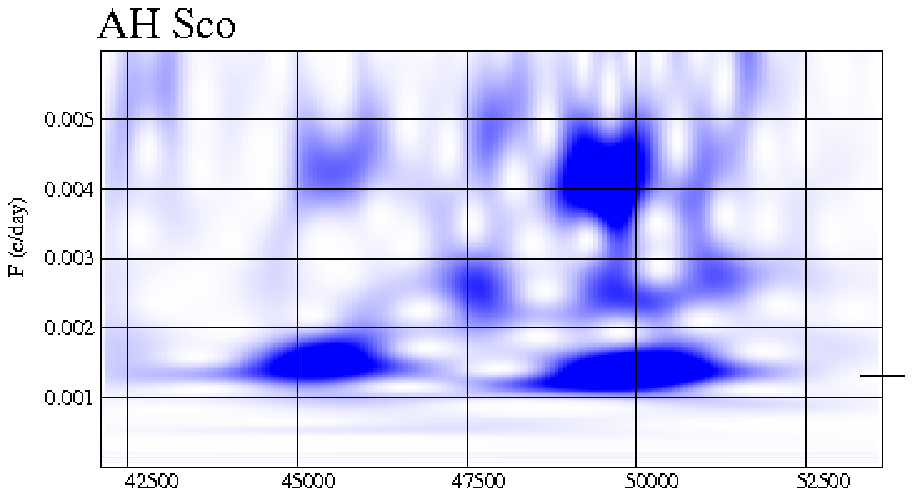}
\includegraphics[width=8cm]{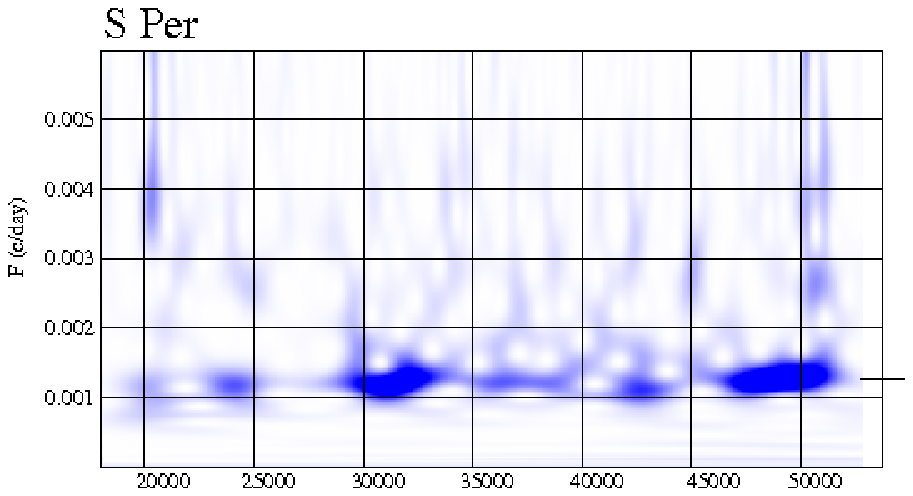}
\includegraphics[width=8cm]{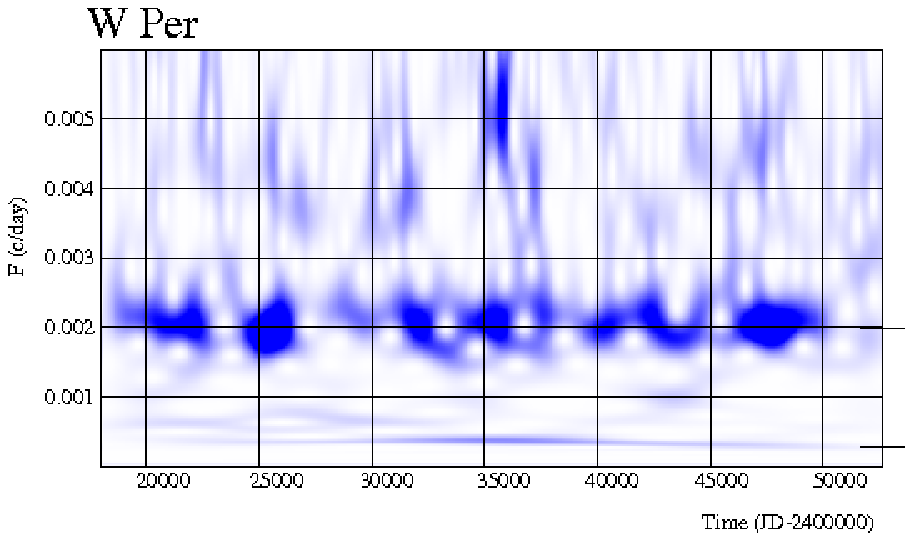}
\end{center}
\caption{Sample wavelet maps. The horizontal ticks on the right hand 
side show the mean frequencies.}
\label{wav}
\end{figure}

To detect periodicities we calculated Fourier spectra of the light curves and then
identified frequencies of power excesses. For this, we used 
Period04 of Lenz \& Breger (2004) to carry out standard iterative sine-wave fitting. In every
iteration, a sine-wave corresponding to the highest peak in the frequency spectrum
was fitted and subtracted from the data. The spectrum was then recalculated using the
residual data. The iterative procedure was stopped when the residual spectrum did not
contain peaks more than three times the noise floor.  We then merged the frequencies in a few well-confined
groups, within which the amplitude-weighted centroid defined a mean ``cycle length'' of
the star. These were later refined using a quantile analysis (see below). In the stars with highest
amplitudes  (e.g. S~Per, VX~Sgr) we found integer multiples of the dominant
frequency, showing the non-sinusoidal shape of the curve; in those cases, the harmonics
were not treated as separate periods. 

To allow direct comparison between the stars, and because the oscillation modes are
resolved, we converted the power (=amplitude$^2$) spectra to power density (power per
frequency resolution bin, expressed in mag$^2$(c/d)$^{-1}$), since the latter is
independent of the length of the observations (Kjeldsen \& Bedding 1995, App. A.1).
Heavily smoothed power density spectra (PDS) were then used to determine the frequency
dependence of the noise (observational as well as astrophysical), which seem to be the
dominant factor in some of the variables. 

Sample power spectra are plotted in Figs.\ \ref{pds1}-\ref{pds2}, where we also show
the spectral window for each star. In high-amplitude stars close to the
ecliptic plane (AH~Sco, VX~Sgr), we see strong yearly aliases, and that is why there
are relatively high peaks in the spectra that remained unmarked. Also, for a few stars
we find relatively strong peaks at a period of exactly 1-yr, whose reality is quite
doubtful. Since these variables have very red colours, visual observations can 
be affected by seasonal poor visibility, when the stars are observed at such high
air-masses that the colour difference between the variable and comparison stars may
lead to differential extinction of a few tenths of a magnitude. We found very similar
1-yr periods in semiregular red giants (Kiss et al.\ 1999), so that periods between 355
and 375 days that were based on a single sharp peak were omitted from further 
analysis. 

As can be seen from the spectral windows in Figs.\ \ref{pds1}-\ref{pds2}, the typical
sampling is excellent and most of the structures in the PDS are real. The closely
spaced peaks and their power distribution is very similar in shape to those of the 
solar-like  oscillators and pulsating red giants with stochastic behaviour (Bedding
2003, Bedding et al.\ 2005). For that reason, hereafter we make an important 
distinction between the instantaneous period of the star, which can be measured from a
shorter subset of the light curve and the mean period, which has the real physical
meaning.  Due to the seemingly irregular nature of the light variations,  there are
certain limitations in assigning ``periods'' to the observations. Assuming that
stochastic excitation and damping occur in these stars, one will always measure 
different period values from datasets that are comparable in length to the mode 
lifetime; however, that does not mean that the physical frequency of that particular 
mode has changed. In that sense we avoid using the term ``period change'', because
what we can measure does not imply any change at all in the period of the pulsation
mode.

Further support to this was given by time-frequency analysis, which can reveal 
time-dependent modulations of the frequency content (e.g. Szatm\'ary, Vink\'o \& G\'al
1994, Foster 1996, Bedding et al.\ 1998,  Szatm\'ary, Kiss \& Bebesi 2003, Templeton,
Mattei \& Willson 2005). We checked several time-frequency distributions for gradual
period evolution using the software package TIFRAN (TIme FRequency ANalysis),
developed by Z. Koll\'ath and Z. Csubry at the Konkoly Observatory, Budapest
(Koll\'ath \& Csubry 2006). In all cases, the wavelet maps clearly showed that close
multiplets in the  power spectra corresponded to a single peak that was subject to a
time-dependent ``jitter'' with no signs of long-term evolution. Sample wavelet maps 
are shown in Fig.\ \ref{wav}, with mean frequencies marked by the horizontal ticks. It
is obvious, for example, that what seems to be two closely spaced peaks in AH~Sco does
in fact reflect a slight shift of of the dominant peak in the middle of the dataset.
Similarly, the broad power excess in W~Per (Fig.\ \ref{pds2}) corresponds to a peak
that is highly variable in time. This behaviour is very typical for most of the stars
and is similar to that of the semiregular red giant variables (e.g. Percy et al 1996,
2003). The fluctuations of the instantaneous frequency usually do not exceed 5--10\%
of the mean, although in extreme cases (like $\alpha$~Ori, TV~Gem) the full width of 
a power excess hump can have relative width of about 20\% in frequency. 

Keeping in mind these time-dependent changes, we determined mean periods as follows. 
We identified probable dominant frequencies from the well-defined humps of power
excesses in the PDS, as described above. Then we selected  the low-
and  high-frequency limits of each hump as the frequencies where the power density
reached the level of the noise.  In this interval the cumulative distribution of the
power was calculated and normalized to 1. The 0.5-quantile, the frequency where the 
normalized distribution crossed the 0.5 value is adopted as the mean frequency of the
hump. Similarly, the width of the power excess was assigned to be the difference 
between 0.166 and 0.834 quantiles, divided by 2.  This way one can assume that at any
given time the instantaneous  period is within the [mean$\pm$width] interval with
1-sigma confidence. 

This method is somewhat subjective with the selection of the initial frequency
intervals. But the quantiles are the most sensitive to the  position of the peak and
this stabilizes the results against the initial  conditions. We tested this with
comparing the results of the above discussed initial setup and another, strongly
different test setup, where one of the initial positions was set twice further from
the other one. The mean frequency was changed only by a few per cent, despite the
obviously wrong initial settings. The width is almost similarly stable with a
variation of 5--20\%. This showed that the uncertainty
of the mean frequency is usually in the order of a few percent, being much smaller
than the natural jitter present in the stars.
In several cases we fitted Lorentzian envelopes to the power excess humps (see details
in Sect.\ 5) to measure the centroid frequency and its damping rate. For every star
with well-defined, regular humps the results were very similar to those of the
quantile analysis. However, for other stars the PDS is quite noisy and the assumption
of Lorentzian power distribution is not true, so that fitting Lorentzians was not
possible for the whole sample.

\section{Results}

\begin{table*}
\begin{centering}
\caption{\label{periods} Periods from this study and the literature. The given
$\pm$range values were calculated from the width of the power distribution in the 
frequency spectra and are dominated by the intrinsic jitter of the stars. 
Numbers in parentheses refer to an upper limit of amplitude (in magnitudes) in the Fourier 
spectrum in those cases where there was no peak with S/N$>$3, while (1/f) refers to
cases where the spectra show a constant rise towards the lowest frequencies with no
discernible peak. Stars in parentheses were rejected as supergiants.}
\begin{tabular}{lrrc|lrrc}
\hline
Star & Period(s)$\pm$range & Period(s) & Source &Star & Period(s)$\pm$range & Period(s) & 
Source \\
  &  this study [d]& literature [d] &  & &  this study [d]& literature 
[d] & \\
\hline
SS~And &  159$\pm$17   & 152.5 & 1 &                            $\alpha$~Her & 124$\pm$5, 500$\pm$50, 1480$\pm$200& 128, long &  10  \\
NO~Aur & (0.05) & ---    & --- &			 	RV~Hya & (0.10) & 116 & 1\\
(UZ~CMa) & 362$\pm$11,38.4$\pm$0.3 & 82.5 & 1 & 	 	W~Ind & 193$\pm$15 & 198.8 & 1\\
       &         & 41   & 2 &				 	(Y~Lyn) & 133$\pm$3, 1240$\pm$50& 110 & 1\\
VY~CMa & 1600$\pm$190 &---& ---   &			 		   &	  & 1190, 133 & 6\\
RT~Car & 201$\pm$25, 448$\pm$146 &---& ---&				  &	 &  110, 1400 &  10\\
BO~Car & (0.08) &---& ---  &				        XY~Lyr & 122 & 120 &  11\\
CK~Car & (1/f) & 525 &	1  &			                $\alpha$~Ori & 388$\pm$30, 2050$\pm$460 & 2200 & 3\\
        &     & 500: & 2&		   		        	 &	 & 2000 & 7\\
CL~Car & 490$\pm$100, 229$\pm$14, 2600$\pm$1000& 513& 1&        	 &	 & 400, 1478 & 8\\
       &    & 952  & 2&					             &   & 2000, 200, 290, 450 & 9\\			       
EV~Car & 276$\pm$26,820$\pm$230 & 347 &	1 &		        S~Per & 813$\pm$60& 822 & 1\\								
       &       & 235 & 2 &				              & 	 &  745, 797, 952, 2857&12\\		       
IX~Car & 408$\pm$50, 4400$\pm$2000 & 400 & 1&		        T~Per & 2500$\pm$460 & 2430 & 1\\					
TZ~Cas & 3100 & --- &	--- &				              &      & 290, 2800& 3\\									
PZ~Cas & 850$\pm$150, 3195$\pm$800& 925 & 1  &		        W~Per & 500$\pm$40, 2900$\pm$300& 485, 2667 & 1\\	       
       &          & 900 & 3 &				               &	  & 467, 3060 & 3\\					
W~Cep & (1/f) & ---& --- &				        RS~Per & 4200$\pm$1500& --- & ---\\						     
ST~Cep & 3300$\pm$1000 & 2050 & 3 &			        SU~Per & 430$\pm$70, 3050$\pm$1200& 533 & 1\\				
$\mu$~Cep & 860$\pm$50, 4400$\pm1060$ & 730, 4400 & 1 &	               &	   & 500 & 3\\  			       
          &           & 4500 & 3&			        XX~Per & 3150$\pm$1000 & 415, 4100 & 1\\					
	  &           & 873, 4700 & 4&			        AD~Per & (1/f) & 362.5 & 1\\						
	  &           &  850       & 9 &             BU~Per & 381$\pm$30, 3600$\pm$1000& 367 & 1\\
	  &           & 840       & 10 &                   &			  &	&  \\ 
(T~Cet) & 298$\pm$3, 161$\pm$3 & 159 & 1 &		               &	   & 365, 2950 & 3\\			     
        &        &  110:, 280: & 11 &              FZ~Per & 368$\pm$13& 184 & 1 \\ 	
AO~Cru & (0.06) & --- &	--- &				        KK~Per & (0.04) & --- & ---\\								
RW~Cyg & 580$\pm$80& 550 & 1 &				        PP~Per & (0.05) & --- & ---\\						
       &     & 586 & 3&					        	PR~Per & (0.05) & --- & ---\\					
AZ~Cyg & 495$\pm$40, 3350$\pm$1100& 459 & 1 &		        VX~Sgr & 754$\pm$56& 732 & 1\\  					
BC~Cyg & 720$\pm$40 & 700 & 1 &				        AH~Sco & 738$\pm$78& 714 & 1\\  					
BI~Cyg & (0.10) & --- &	---&				        $\alpha$~Sco & 1650$\pm$640 & 1733 & 3\\				
TV~Gem & 426$\pm$45, 2550$\pm$680 & 400, 2248 & 5&	        	     &  	    &  350 &  9 \\				
       &           & 182    & 3&			        CE~Tau & 1300$\pm$100& 165 & 1\\			       
WY~Gem & 353$\pm$24 & --- & --- & 	                               &       & 140-165, $>$730 & 8\\
BU~Gem & 2450$\pm$750 & --- & --- &		                	&	& 272, 1200: &  9\\				 
(IS~Gem) & (0.02) & --- & --- &			                W~Tri & 107$\pm$6, 590$\pm$170& 108 & 1\\
\hline
\end{tabular}
\\
\vskip3mm
Sources: 1 - GCVS (+ notes); 2 - ASAS (Pojmanski 2002); 3 - Stothers \& 
Leung (1971); 4 - Mantegazza (1982); 5 - Wasatonic, Guinan \& Engle (2005); 
6 - Szatm\'ary \& Vink\'o (1992); 7 - Goldberg (1984); 8 - Wasatonic \& 
Guinan (1998); 9 - Percy et al.\ (1996); 10 - Percy, Wilson \& Henry (2001); 
11 - Percy et al.\ (2001); 12 - Chipps, Stencel \& Mattei (2004)
\end{centering}
\end{table*}

We present the derived periods in Table\ \ref{periods}. In total, we determined 56
periods for 37 stars; in  6 cases ours is the
first period determination in the literature. We found published periods for 31 stars,
but except for a few well-studies variables (e.g. $\mu$~Cep, $\alpha$~Ori, S~Per),
most of them were quite neglected in the past three decades. Nevertheless, the overall
agreement with the catalogued  periods is good: for the shortest period stars (e.g.
SS~And, T~Cet, Y~Lyn, W~Tri) our findings are in perfect agreement with the GCVS or
other, more recent studies (with the differences staying below 2--3\%). In a few cases
we could not infer any clear coherent signal from the curves, so we show the upper
limits on the semi-amplitude in parentheses. 

\begin{figure}
\begin{center}  
\leavevmode
\includegraphics[width=8.5cm]{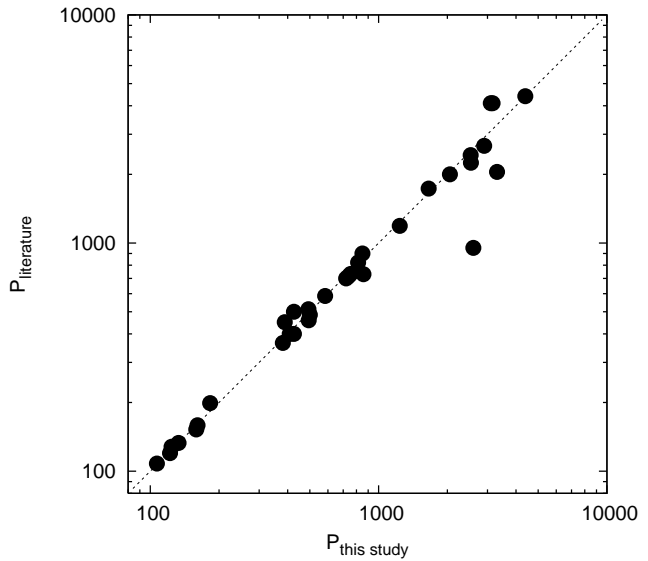}
\end{center}
\caption{A comparison of 33 periods for 28 stars.}
\label{percomp}
\end{figure}

In Fig.\ \ref{percomp} we compare our periods with the literature.  For this
comparison we selected those periods from the literature that were definitely
corresponding to ours. There are three significant outliers, namely CL~Car at 2600 d
(with 952 d from ASAS project), ST~Cep at 3300 d (with 2050 d from Stothers \& Leung
1971) and XX~Per at 3100 d (with 4100 d also from  Stothers \& Leung 1971). In case of
CL~Car, we checked the ASAS observations of the star and found that the 952 d period
is an error. For ST~Cep and XX~Per our data have significantly longer time-span, so
that the newly determined values are likely to be more accurate.  

We have attempted to confirm the red supergiant status of stars in our sample. The
minimum absolute bolometric magnitude for red supergiants is about $-5$ mag (Meynet \&
Maeder 2003), which can be translated to $M_{\rm K}^{\rm min}\approx-8$ mag using the
relation $m_{\rm bol}\approx m_{\rm K}+3$ that connects extinction-corrected
bolometric and $K$-band magnitudes (Josselin et al.\ 2000). In our sample, bolometric
magnitudes were recently determined for 18 stars by Levesque et al.\ (2005) and of
these only ST~Cep is located near the minimum luminosity, with $M_{\rm bol}=-5.48$ mag.
A few stars have useful Hipparcos parallaxes ($\alpha$~Ori, $\alpha$~Sco,
$\alpha$~Her, CE~Tau), and all are at least 1 mag brighter than $M_{\rm K}=-8$ mag. A
particularly interesting star is $\alpha$~Her because its period (125 d) is very 
short for a supergiant star. For instance, Jura
\& Kleinmann (1990) rejected all red supergiant candidates with periods less than 150
d, although it was clearly established from long-period variables in the Large
Magellanic Cloud that the least luminous red supergiants can have periods between 100
and 150 d (Wood, Bessell \& Fox 1983). For $\alpha$~Her, $K=-3.70$ mag (Richichi \&
Percheron 2002) and $\pi=8.5\pm2.8$ mas (ESA 1997), implying $M_{\rm K}=-9.0\pm0.7$ mag,
so our conclusion is that shorter periods ($100-150$ d) can indeed exist in RSGs.

For four stars, whose names are shown in parentheses in Table\ \ref{periods}, 
the RSG class is somewhat doubtful.
For UZ~CMa, both the AAVSO light curve and the ASAS $V$-band observations
infer a periodicity around 40 days; the ASAS CCD-$V$ light curve is very typical of a
short period red giant semiregular variable, presumably on the AGB, which is also
supported by the luminosity class II.  IS~Gem, although
classified as SRc, has too early a spectral type (K3II) 
and its $K$-band absolute magnitude is only $-2.10$ mag, being consistent 
with the luminosity class.  For T~Cet, the periods and their ratio of 1.87 are fairly
typical for semiregular AGB stars (Kiss et al.\ 1999), and its Hipparcos parallax and
2MASS $K$ magnitude imply M$_{\rm K}=-7.7\pm0.4$ mag, which is too faint. Y~Lyn is a
very typical semiregular variable with a long-secondary period, which is a well-known
but still poorly understood phenomenon in AGB stars (Wood, Olivier \&
Kawaler  2004). Moreover, its $K$-band absolute magnitude from the Hipparcos parallax
and 2MASS $K$-magnitude is only $-7.7\pm0.7$ mag. For these reasons, hereafter we
exclude these four stars from further analysis.

\section{Discussion of the multiperiodic nature}

According to the theoretical predictions, multiple periodicity may arise from multimode
pulsations. Radial oscillations have long been predicted by model calculations, starting
from the pioneering work of Stothers (1969). Two extensive investigations on pulsation
properties of RSGs were recently published by  Heger et al.\ (1997) and Guo \& Li (2002).
Although they treated convection in a different way (Heger et al.\ adopted the
Ledoux criterion and treated semiconvection according to Langer, Sugimoto \& Fricke
(1983); Guo \& Li used the mixing-length theory of B\"ohm-Vitense (1958) and adopted the
Schwarzschild criterion to determine the boundaries of convection and semiconvection
zones), both studies confirmed earlier results and predicted excitation of the
fundamental, first and possibly second overtone modes. The fundamental mode's growth rate
always exceeded that of the overtone modes and was found to increase with luminosity for
a given mass.  On the other hand, studies of convection in the envelopes of red giants
and supergiants showed that the dominant convective elements can be comparable in size to
the stellar radius, which could explain both the observed irregular variations
(Schwarzschild 1975,  Antia et al.\ 1984) and the hotspots on late-type supergiants,
detected using interferometric techniques (Tuthill, Haniff, \& Baldwin 1997). A similar
mechanism was proposed by Stothers \& Leung (1971) for explaining the longer periods of
RSG variables, arguing that the similarity of those periods and the time scale of
convective turnover suggests a link between the two phenomena. Interestingly, periods
that are too long were also found in other luminous stars. Maeder (1980) calculated
empirical pulsation constants of blue-yellow  supergiants, which were found to be
systematically larger than the theoretical Q value for the fundamental mode of radial
oscillation. Maeder (1980) suggested  that those long periods can be due to non-radial
oscillations of gravity modes. The same mechanism was proposed by Wood, Olivier \&
Kawaler (2004) as a possible explanation of the long-secondary periods of AGB stars. 

\subsection{Period--luminosity relations}

\begin{figure}
\begin{center}  
\leavevmode
\includegraphics[width=8.5cm]{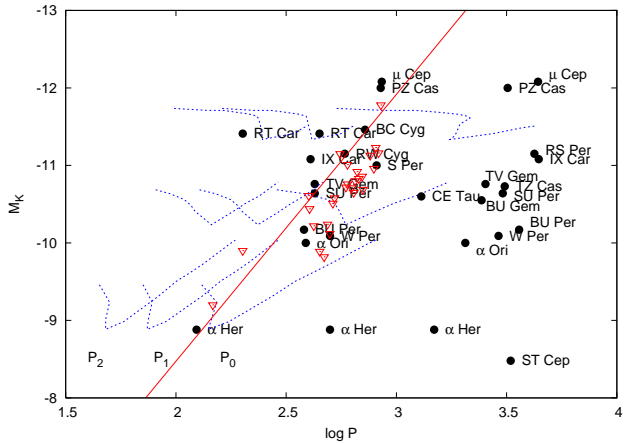}
\end{center}
\caption{Period--absolute magnitude relations for 30 periods of 18 stars. Triangles 
refer to RSG variables in the LMC, with periods and photometry taken from 
Wood, Bessell \& Fox (1983). Solid circles and asterisks were divided at 
P=1000 d. The dotted lines are based on model calculations by Guo \& Li (2002) for
solar metallicity, while the solid lines shows the best-fit period--$K$ magnitude
relation.}
\label{pl}
\end{figure}

A significant fraction of our sample can be characterized by two dominant periods, one
of a few hundred days and one of 1500--2000 days. The strong similarity
that we found in many cases argues for the reality of these long periods. Some of the
light curves are more than twice as long as those analysed by Stother \& Leung (1971)
and the fact that we derive very similar values for the long periods indicates the
reliability  of the results. To reveal deeper insights into the nature of the multiply
periodic variations, we studied the
period--luminosity distribution for all stars having a useful estimate of 
luminosity. 

To construct the period--$K$-band absolute magnitude diagram, we took the following
steps: {\it (i)} $M_{\rm bol}$ values from Levesque et al.\ (2005) were converted to
$M_{\rm K}$ using the Josselin et al.\ (2000) relation; {\it (ii)} for $\alpha$~Her,
$\alpha$~Ori and CE~Tau we used Hipparcos parallaxes and K-band data to calculate
$M_{\rm K}$ directly; {\it (iii)} for comparison we added RSGs in the Large Magellanic
Cloud, taken from Wood, Bessell \& Fox (1983). The resulting P--L diagram is shown in
Fig.\ \ref{pl}, where the lines represent fundamental, first overtone and second
overtone modes for models of solar metallicity by Guo \& Li (2002). The help
distinguish between the two sequences, periods longer than 1000 d were plotted with
different symbols.

The shorter periods are well matched by the fundamental and first overtone modes of the
models. Moreover, there is perfect agreement with the LMC red supergiant sample, too.
$\mu$~Cep and PZ~Cas are above the luminosity range of the models and it is possible
that in those cases the longer period is the fundamental mode. RT~Car and maybe IX~Car
seem to be too luminous  for fundamental pulsation, but otherwise models strongly
favour fundamental or first overtone modes for the shorter periods. It is worth noting
that $\alpha$~Her has an almost identical counterpart in the LMC, which provides a
retrospective confirmation of the reality of the 125 d period.

Excluding the short period of RT~Car, we fitted the following period--$K$ magnitude
relation to the solid dots in Fig.\ \ref{pl}:

$M_K =(-3.44\pm0.6) \log P + (-1.6\pm1.6)$

\noindent with 0.46 mag rms, which is about the typical uncertainty of individual
absolute magnitudes. The slope of this relation is very similar to the
slope of the Mira P--L relation in the Large Magellanic Cloud ($-3.52\pm0.03$, Ita et
al.\ 2004). The agreement shows the similarity of pulsations in red giant and
red supergiant stars. However, the zero points are very different, as the Ita et al.
relation for fundamental mode AGB stars has a zero point of +1.54$\pm$0.08 mag (with
$\mu_{\rm LMC}=18.50$).

The long secondary periods (LSPs) remain beyond the limits of the models, and both the
updated models and the refined physical parameters of the stars keep the original
conclusion on the peculiar nature of LSPs by Stothers \& Leung (1971) unchanged. They
cannot be explained by radial pulsations, because the period of the fundamental mode
is the longest possible for that kind of oscillation. Metallicity effects also cannot
explain the LSPs. Looking at the positions, for instance, $\alpha$~Her or
$\alpha$~Ori, their absolute magnitudes are 1--2 mags fainter than those of the solar
metallicity models. According to Guo \& Li (2002), $\delta M_{\rm bol}\sim0.83\delta
\log Z$, which means that unphysically large metallicities (10--100 times solar) would
be needed to account  for the low luminosities. A possibility is that heavy
circumstellar extinction in $K$-band makes the stars fainter. Massey et al.\ (2005)
indeed found many magnitudes of circumstellar $V$-band extinction in a number of
galactic  RSGs, which showed that the effect may not be negligible (for example, in
Fig.\ \ref{pl} $\alpha$~Her, $\alpha$~Ori and CE~Tau were not corrected for this).
Furthermore, it is known that the reddest Mira stars in the Small Magellanic Cloud
have $K$-band magnitudes that are fainter by 1--2 mag than predicted by the Mira P--L
relation (Kiss \& Bedding 2004), which is similar to what we see here for the red
supergiants. On the other hand, the very good agreement for the shorter periods
between the galactic and the LMC samples argues against this explanation (see, e.g.,
$\alpha$~Her and the closest LMC point in Fig.\ \ref{pl}). Also, the 4--5 mags extra
optical extinction that has been found in cluster RSGs by Massey et al.
(2005) could hardly explain the 1--2 mags extra extinction in the $K$-band, unless the
extinction law of the  circumstellar matter is extremely different of the ``standard''
one  ($A_K/A_V=0.112$, Schlegel, Finkbeiner \& Davis 1998). Therefore, the LSP P--L
sequence is a separate entity, whose origins needs further investigation (possibly
together with the AGB LSP phenomenon).

\begin{figure}
\begin{center}  
\leavevmode
\includegraphics[width=8.5cm]{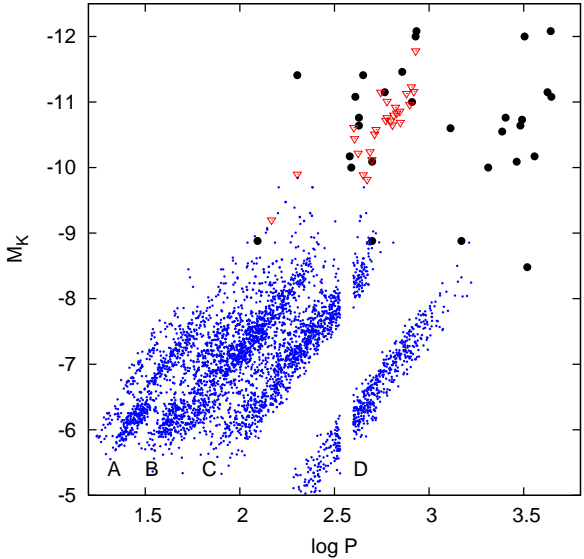}
\end{center}
\caption{Period--$K$-magnitude relations for the galactic sample (black circles),
RSG variables in the LMC (red triangles) and red giant variables from the MACHO database
(blue dots, data taken from Derekas et al.\ 2006, assuming $\mu_{\rm LMC}=18.50$). Labels
A, B, C and D were adopted from Wood (2000).}
\label{rsg+lmc}
\end{figure}

In Fig.\ \ref{rsg+lmc} we compare the period--$K$-magnitude relations for RSGs in our
galactic sample (large black dots) and in the LMC (red triangles) with red-giant
variables in the LMC (small blue dots), as observed by the MACHO project (data taken
from Derekas et al.\ 2006).  The MACHO sample comprises low-mass stars on the RGB and
AGB.  There is some similarity between the P--L relations of the supergiants and the
less luminous RGB and AGB stars, although the lack of precise distance estimates for
our sample of RSGs makes a detailed comparison difficult.  Also, we expect the RSG
sequences to be broad because these stars cover a large range of stellar masses. 
Nevertheless, it does appear that the shorter periods of the RSG stars mostly align
better with Sequence B of the low-mass stars rather than with Sequence C, which would
imply pulsation in the first-overtone rather than the fundamental.  It would clearly be
valuable to have long time series for a large sample of LMC stars, which could be
obtained by analysing the photographic plate archives.

\subsection{Pulsation constants}

\begin{figure}
\begin{center}  
\leavevmode
\includegraphics[width=8cm]{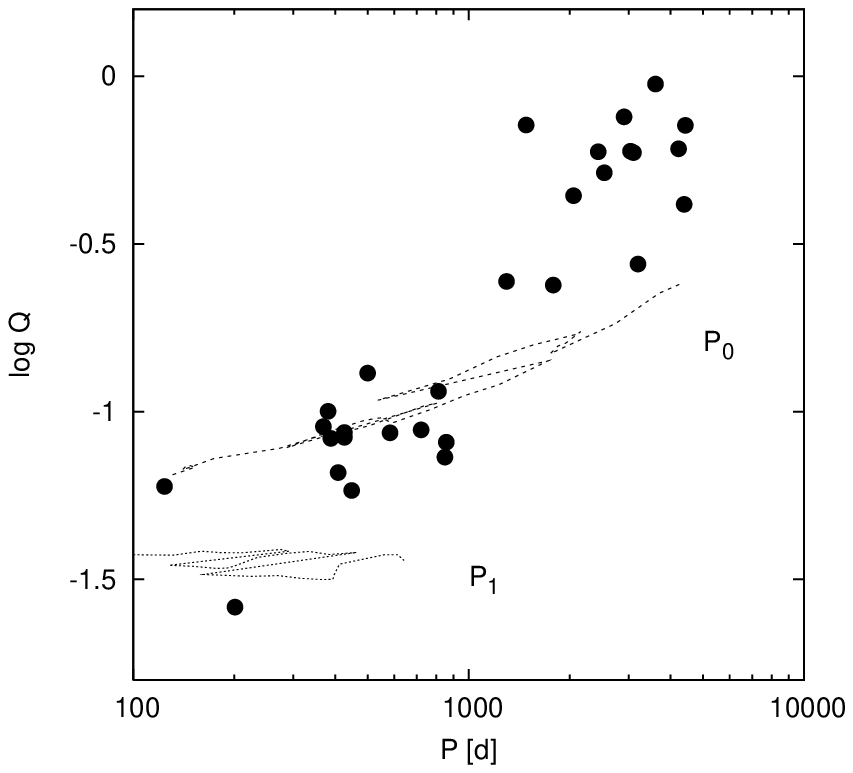}
\includegraphics[width=8cm]{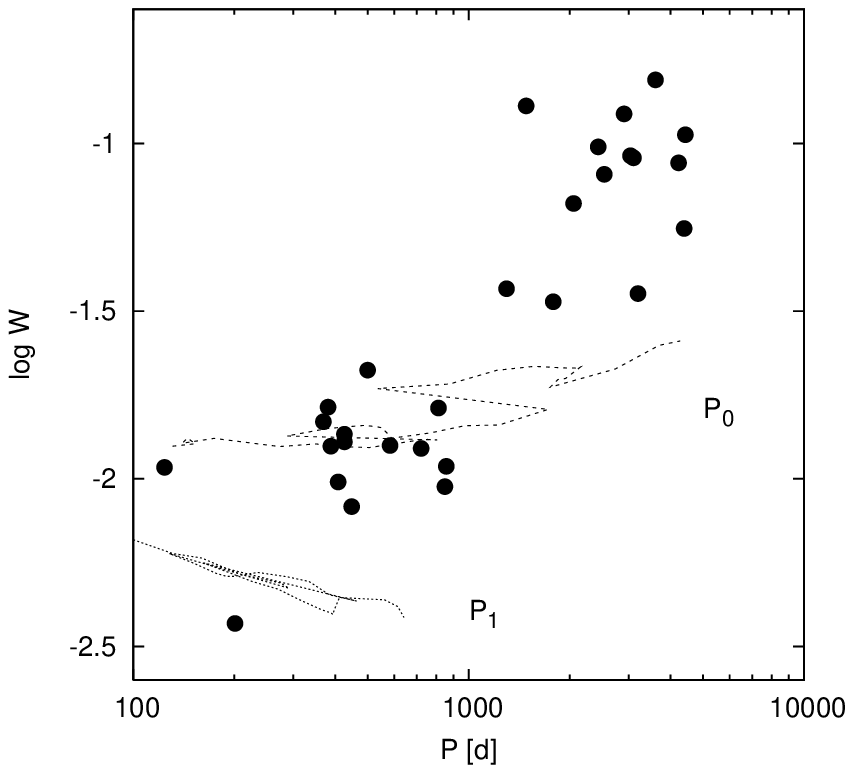}
\end{center}
\caption{Pulsation constant $Q$ and $W$. The dashed and dotted lines show the 
fundamental and first overtone models of Guo \& Li (2002). The downward outlier 
at P = 200 d is RT~Car.}
\label{modelq}
\end{figure}

Further interesting details are revealed by examining the pulsation constants
$Q=P(M/M_\odot)^{1/2}(R/R_\odot)^{-3/2}$ and $W=P(M/M_\odot)(R/R_\odot)^{-2}$. The
first is the classical period-density relation, while the second is the natural form
of the pulsation constant if the oscillations are confined to the upper layers of the
envelope (Gough, Ostriker \& Stobie 1965). Stothers (1972) showed that for pulsation
models of massive red supergiants, $W_0$ of the fundamental mode was more constant
than $Q_0$. We have calculated both quantities for the stars in Fig.\ \ref{pl} as
follows. Radius values were either taken from Levesque et al.\ (2005) or calculated 
from the parallax-based luminosity ($\alpha$~Her and CE~Tau) and temperature
($\alpha$~Her: Levesque et al.\ 2005; CE~Tau: Wasatonic \& Guinan 1998). Masses were
estimated from the approximate relation $\log (M/M_\odot)=0.50-0.10 M_{\rm bol}$,
which comes from evolutionary calculations (Levesque et al.\ 2005); for this, $M_{\rm
bol}$ values were taken from Levesque et al.\ (2005) or calculated from $M_{\rm K}$.  

We plot the resulting pulsation constants $Q$ and $W$ as function of period in  Fig.\
\ref{modelq}. In addition to the empirical values, we also show fundamental- and first
overtone-mode models of Guo \& Li (2002). Based on this diagram we draw several 
conclusions. Firstly, looking at the model calculations, $W$ of the fundamental mode 
is indeed a better constant than $Q$, in accordance with the theoretical expectations,
while the first overtone-mode models show an opposite behaviour. 
Secondly, all
periods less than 1000 d agree with fundamental pulsation for both $Q$ and $W$ (except
for RT~Car). Thirdly, there is a remarkable  feature of the pulsation constants: the
mean value of $Q$ for the shorter periods and  the mean value of $W$ for the long
periods are practically equal: $\langle Q \rangle=0.085\pm0.02$ and $\langle W
\rangle=0.082\pm0.03$. Although this might be coincidence, it may suggest that the
fundamental period and the long secondary period are intimately connected via the
relation $P_0/P_{\rm LSP}=(M/M_\odot)^{1/2}(R/R_\odot)^{-1/2}$, which could be used as
test for future models of the LSPs.

One application of the two pulsation constants could be checking the
consistency of a given set of physical parameters for a star with the observed
periodicity. A notoriously ambiguous  case is VY~CMa, for which vastly different
radius estimates can be found in the literature, ranging from 600 $R_\odot$ up to
2,800 R$_\odot$ (Massey, Levesque \& Plez 2006 and references therein). For example,
Monnier et al.\ (1999) adopted $M\approx 25$ M$_\odot$ and $R\approx 2,000$ R$_\odot$,
while Massey, Levesque \& Plez (2006), using the new temperature scale of Levesque et
al.\ (2005), arrived to $M\approx 15$ M$_\odot$ and $R\approx 600$ R$_\odot$. Using the
mean cycle length of 1,600 d, the high-mass/large radius parameter set results in
$Q=0.089$ and $W=0.01$, both being consistent with fundamental mode pulsation. For the
low-mass/small radius set, the two constants are $Q=0.42$ and $W=0.066$, which could
be acceptable if the 1600 d period referred to a long secondary period. At this stage,
unfortunately, the periodicity does not help solve the problem, but since no other 
RSG has a fundamental-mode period greater than 1000 d, we have a
slight preference for the low-mass/small radius set and the LSP interpretation of the
1,600 d periodicity. 

\section{Evidence for stochastic oscillation and $1/f$ noise}

\subsection{Lorentzian envelopes in the power spectra}

The AAVSO observations span many decades and sometimes almost a century, thus we are now
in a better position to understand the {\it irregularity} of RSG variability than
previous researchers. Not much effort was put into this direction in the past:
irregularity was taken as a general description of non-periodic brightness fluctuation in
red giants and supergiants. For semiregular red giants the nature of irregularities was
addressed by Lebzelter, Kiss \& Hinkle (2000), who compared simultaneous light and
velocity variations in a sample of late-type semiregular variables (of the GCVS types SRa
and SRb) and concluded that the observed variability is most likely a combination of
pulsations and additional irregularity introduced by, e.g., large convective cells. The
latter phenomenon is theoretically expected  (Schwarzschild 1975, Antia et al.\ 1984),
although not much is known on the time-dependent behaviour of the integrated flux
variations that arise from the huge convective cells. Percy et al.\ (2003) found
growth/decay timescales of 1--5 years in a sample of small-amplitude pulsating red giant
stars, which they interpreted as the natural growth (or decay) times for the pulsation
modes. Recently, Bedding et al.\ (2005) discussed the possibility of solar-like
excitation of the semiregular variable L$_2$~Pup. In that star the the power distribution
closely resembles that of a stochastically excited damped oscillator and is  strikingly similar to close-up
views of individual peaks in the power spectrum of  solar oscillations: there is a single
mode in the spectrum which is resolved into multiple peaks under a well-defined
Lorentzian envelope. The envelope's width gives the damping time (or mode lifetime),
which is one of the main characteristics  of a damped oscillator. Bedding et al.\
(2005) argued that the close similarity may imply solar-like excitation of oscillations,
presumably driven by convection. Another interpretation could be that the cavity of
oscillations is changing stochastically due  to the convective motions, which affects the
regularity of the pulsations.  

\begin{figure}
\begin{center}  
\leavevmode
\includegraphics[width=8cm]{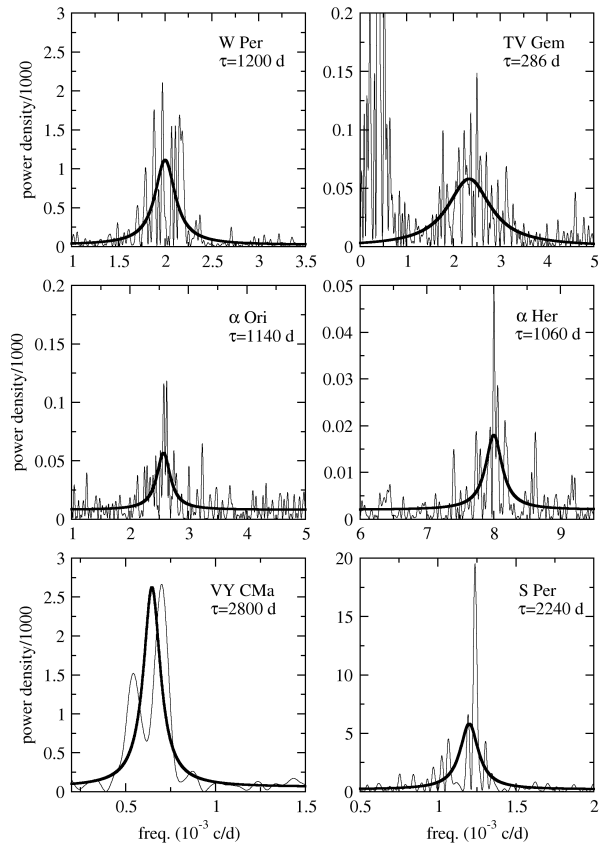}
\end{center}
\caption{Lorentzian fits (thick lines) of the power spectra (thin lines).}
\label{lorentz}
\end{figure}

We find the very same structures in the power spectra of several RSGs: a well-defined
Lorentzian envelope under which the power is split into a series of narrow peaks.
Probably the best example is W~Per (Fig.\ \ref{pds2}), where the window function is
practically free of any kind of alias structure, so that no false structure due to
poor sampling arises in the power spectrum. Other convincing  examples include
$\alpha$~Ori, TV~Gem, S~Per, AH~Sco, VY~CMa and $\alpha$~Her, whereas in a few stars
the power is so spread over a large range of frequencies that no regular envelope can
be traced in the spectrum.

Adopting the same approach as Bedding et al.\ (2005), i.e. assuming a
stochastically excited damped oscillator, we have fitted Lorentzian profiles to the
power spectra assuming $\chi^2$ statistics with two degrees of
freedom. This is based on a maximum-likelihood fit, assuming an exponential
distribution of the noise (Anderson, Duvall \& Jefferies 1990, Toutain \& Fr\"ohlich
1992). The fit gives the centroid frequency and the half-width at half power
$\Gamma$, which can be converted to mode lifetime via $\tau=(2\pi\Gamma)^{-1}$.    

\begin{figure*}
\begin{center}  
\leavevmode
\includegraphics[width=\textwidth]{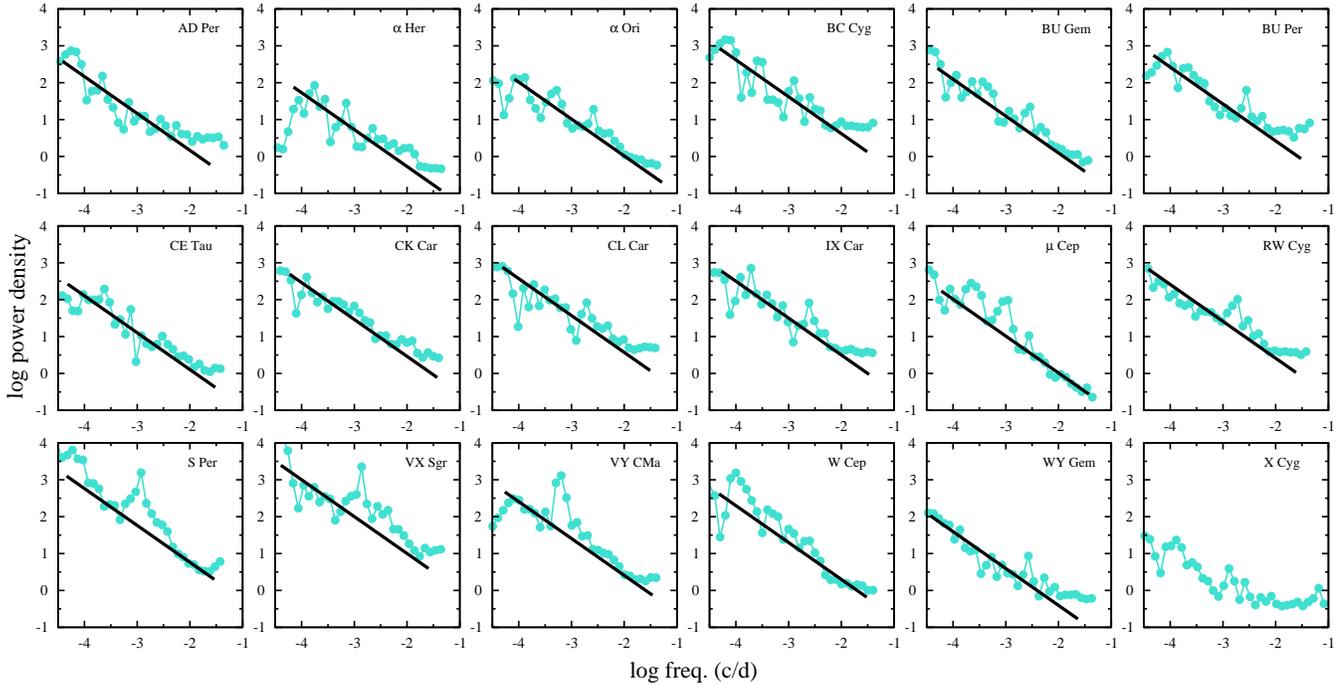}
\end{center}
\caption{Smoothed power density spectra in log-log representation. The thick line shows
the best-fit $1/f$ noise. For comparison, the lower right panel
shows the log-log PDS of the regular Cepheid variable X~Cyg.}
\label{1perf}
\end{figure*}

We show the clearest examples of the Lorentzian fits in Fig.\ \ref{lorentz}. We have
looked for correlations between mode lifetime, luminosity and length of the long
secondary periods. Generally, the mode lifetime is typically 3--4 times the pulsation
cycle, with notable exceptions of TV~Gem (less than 1) and $\alpha$~Her (about 8).
There might be a slight correlation between the long secondary period and the mode
lifetime, but the small number of multiperiodic stars with well-defined Lorentzian
fits prevented a firm conclusion.

\subsection{$1/f$ noise}

We also investigated the nature of irregularities through the shape  of the noise
level in the power spectra. The analysis of fluctuation power spectra density is a
common tool in studies of unpredictable and seemingly aperiodic variability, i.e.
noise that arises from a stochastic process. In this context, the noise is intrinsic
to the source and not a result of measurement errors (such as Poisson noise). In
astrophysics, X-ray light curves of active galaxies and interacting binaries have been
a major inspiration of such studies (e.g. Vaughan et al.\ 2003 and references therein).
Of particular interest are noise series whose power spectra are inverse power
functions of frequency, the so-called $1/f^\alpha$ noises, for which examples have
been found in a wide range of natural phenomena (Press 1978). Bak, Tang \& Wiesenfeld
(1987) showed that dynamical systems with spatial degrees of freedom naturally evolve
into a self-organized point. $1/f$ noise in these systems simply reflects the dynamics
of a self-organized critical state of minimally stable clusters of all length scales,
which in turn generates fluctuations on all time scales. Specifically, turbulence is a
phenomenon, for which self-similar scaling  is believed to occur both in time and
space. Therefore, combining the Bak, Tang \& Wiesenfeld (1987) theory with the
Schwarzschild (1975) mechanism of producing irregular variability via large convective
cells, one expects a strong $1/f$ noise component in the power spectra of RSG
brightness fluctuatios, similarly to solar granulation background (e.g. Rabello-Soares
et al.\ 1997). 

That is exactly what we find for the vast majority of the sample. In Fig.\ \ref{1perf} we
plot smoothed  power density spectra in log-log representation. These were calculated
from the initial spectra after transforming to the log-log scale and then binning them
with a stepsize of 0.1 dex. Besides 17 RSGs, we also show the averaged power density
spectrum of the AAVSO observations of the classical Cepheid variable X~Cygni, which
varies between $m_{\rm vis}=5.9-6.9$ mag with a period of 16.38 d. This star, being a
strictly regular pulsating variable, served as a test object for the power distribution
of visual observational errors. As indicated by the lower right panel of Fig.\
\ref{1perf}, the power distribution is much flatter for almost three orders of 
magnitude, with only a slight rise towards the lowest frequencies. This  is likely to
arise from the fact that not many AAVSO observers are active for more than 10,000 days,
so there is a fluctuation of observers on time-scales of longer than a few 1000 days. It
is also apparent that the mean power in X~Cyg is much lower than in any of the RSGs,
scattering between log~PD = 0--1. Its brightness range is roughly in the middle of the
sample (see Table\ \ref{sample}), the time-span is about 20,000 days, so that the mean
noise level should be representative of the typical AAVSO light curve in our set.

The RSG spectra look remarkably similar to each other. There is a
roughly linear power increase towards the low-frequency end, which has very similar
slope in almost all cases. Generally, we do not detect any flattening at the lowest 
frequencies, except for $f<1/T_{\rm obs}$ ($T_{\rm obs}$ is the time-span of the  data),
which suggests that the observations are too short to extend over the whole range of
time-scales of fluctuations that is present in these stars (with $\alpha$~Her being an
exception). Formal linear fits of the spectra resulted in slopes ranging from 0.8 to
1.2, but none was significantly different of 1. Furthermore, distinct features in the
spectra  (like the peaks of the periods) made it difficult to fit the overall slope of
the ``noise continuum'' accurately, and because of that we fixed the slopes at 1 and
fitted the zero-points only, excluding the frequency ranges of the higher power
concentrations. The results are indicated by the thick black lines in Fig.\
\ref{1perf}. 

Based on the close similarity of the log-log power spectra we conclude that there is a
universal frequency scaling behaviour in the brightness fluctuations of red supergiant
stars that fits very well the expectations for background noise from convection. It is
not surprising that period determination is so difficult for these variables: the
longer we observe, the more power will be detected in the low-frequency range,
suggesting the presence of more and more ``periods'' that are comparable to the full
length of the data. Periods identified with the fundamental mode are
definitely real -- although strongly affected by the high-noise in the systems. We
think most of the long-secondary periods are also real, because we see a consistent
picture in many different stars. However, random peaks at the lowest frequencies
naturally develop due to the strong $1/f$ noise signal and it would be misleading to
interpret each peaks as periods. Particularly good examples are AD~Per and W~Cep, of
which the latter has variations up to 2 mags in visual but the power follows a
well-defined $1/f^\alpha$ distribution (with a slope that may be a bit larger than 1).
The whole set of phenomena is strikingly similar to what is observed in the Sun and
other solar-like oscillators. The photometric granulation noise in a main-sequence 
solar-like oscillator has a time-scale of minutes and micromagnitude amplitudes. At
the other end of the spectrum we see these pulsating red supergiants with time-scales
of years and noise amplitudes reaching tenths of a magnitude; the underlying physical
mechanisms seem to be the same all across the Hertzsprung--Russell-diagram. 

Finally, there is interesting correlation between the zeropoints of the fitted lines
and the light curve amplitudes. The two highest amplitude stars, S~Per and VX~Sgr,
have the highest noise levels, reaching one to two orders of magnitudes higher at
$\log f \leq -4$ than any other variable. This shows they are fundamentally more
dynamic in pulsation and noise generation, which reminds us the predicted
``superwind'' phase of RSGs just preceding the supernova explosion. Pulsation model
calculations by Heger et al.\ (1997) have shown that very large pulsation periods,
amplitudes and mass-loss rates may be expected to occur at and beyond central helium
exhaustion over the time-scale of the last few 10$^4$ years. The physical reason for
this is the resonant character of pulsations when the pulsation period and the
Kelvin-Helmholtz time-scale of the pulsating envelope evolve into the same order of
magnitude. A similar result was found by Bono \& Panagia (2000), whose pulsation
models showed larger amplitude and more irregular variations for lower values of
T$_{\rm eff}$, which turned out to be the main governer of the pulsational behaviour.
As Lekht et al.\ (2005) noted for S~Per, overall dimming of the star after a period of
stronger oscillations may be due to subsequent enhanced mass-loss and ejection of a
dust shell that screens the stellar radiation. This is clearly seen for S~Per and
VX~Sgr, and maybe in AH~Sco and VY~CMa, giving supporting evidence for variations of
the pulsation driven mass-loss on a time-scale of $\sim$20 years.

\section{Conclusions}

Red supergiants as variable stars have long been known for their semiregular
brightness fluctuations. Using the extremely valuable database of visual observations
of the AAVSO, we were able to study the main characteristics of their regular and
irregular variations. From a detailed analysis of power spectra and
time-frequency distributions, this paper discussed the properties of pulsations and
their physical implications.

The sample contains several types of light variations. A few stars (S~Per, VX~Sgr,
AH~Sco) have very large peak-to-peak amplitudes that may reach up to 4 magnitudes in
the visual. These objects show both the most coherent periodic signals and the highest
level of $1/f$ noise in the power spectrum, which might be a sign of a ``superwind''
phase of RSGs that precedes the supernova explosion. We argue that multiple periods
found by other studies were sometimes artifacts caused by the extreme noise levels of
the systems. A more common type of variability is characterized by two distinct
periods, one of a few hundred days and one of a few thousand days. The archetypes of
these stars are $\alpha$~Ori and TV~Gem, both having a period around 400 days and
another around 2000 days. The shorter periods can be identified with the radial
fundamental or low-order overtone modes of pulsation, while the longer one is very
similar to the Long Secondary Periods of AGB stars, whose origin is not known yet, but
could be due to binarity, magnetic activity or non-radial $g$-modes (Wood et al.\ 2004).
Besides fundamental pulsation, we also see evidence for first and possibly second
overtone modes. Finally, in a few stars we cannot infer any periodicity at all (e.g.
W~Cep, AD~Per) and these are the truly irregular variables dominated by the $1/f$ noise
in the power spectrum.

The period jitter of the pulsation modes produces in the power spectrum a well-defined
Lorentzian envelope. Interpreting this as evidence for stochastically excited and
damped oscillations, we measured the mode-lifetime (or damping rate) in red
supergiants for the first time. In most stars it is several times longer than the
period of pulsations. Since the damping rate depends on the stellar structure and
convection properties in a complex way, with many weakly constrained theoretical
parameters (e.g. Balmforth 1992) and, moreover, currently there are no theoretical
calculations directly applicable to red supergiants, it is impossible to qualify the
agreement with theoretical expectations. However, the strong $1/f$ noise component
that seems to be ubiquitous in the whole sample, strongly favours the Schwarzschild
(1975) mechanism of producing random brightness variations with huge convection
cells, analogous to the granulation background seen in the Sun.

Finally, stochastic oscillations discussed in this paper may offer an 
explanation for the seemingly random behaviour of less luminous red giants stars. For
example, random cycle-to-cycle fluctuations of the periods of Mira stars (e.g. 
Eddington \& Plakidis 1929, Percy \& Colivas 1999) are likely to be analogous 
to the period jitter we found here. It would be helpful to have theoretical models that
are specifically designed for these phenomena, because the extensive records of
homogeneous visual observations allow us measuring noise properties of red giant 
and supergiant stars quite accurately. In principle, internal physics could be 
probed through these accurate measurements provided that realistic models are 
calculated.     

\section*{Acknowledgments} 

This work has been supported by a University of Sydney Postdoctoral Research
Fellowship, the Australian Research Council, the Hungarian OTKA Grant 
\#T042509 and the Magyary Zolt\'an Public Foundation for Higher Education. 
We sincerely thank variable star observers of the AAVSO whose
dedicated observations over many decades made this study possible. LLK also
thanks the kind hospitality of Dr.\ Arne Henden, the Director of the AAVSO, and
all the staff members at the AAVSO Headquarter (Cambridge, MA) during his visit
in early 2006. The NASA ADS Abstract Service was used to access data and
references.

\end{document}